\documentclass[a4paper, 11pt]{article}
\pdfoutput=1
\usepackage{jcappub}
\usepackage[utf8]{inputenc}

\usepackage{url}
\usepackage{graphicx} 
\usepackage{float}
\usepackage{subcaption}  
\usepackage{hyperref}   

\usepackage{fancyhdr} 

\usepackage{amsmath}
\usepackage{mathtools}
\usepackage{amssymb}
\usepackage{cancel}
\usepackage{units}
\usepackage{bigints}
\usepackage{array}
\usepackage{multirow}

\numberwithin{equation}{section}

\DeclareMathOperator{\arctanh}{arctanh}

\newcommand{\al}{\alpha}

\newcommand{\xmax}{x_{\rm max}} 

\definecolor{darkgreen}{rgb}{0,0.5,0}

\title{Dark energy from $\alpha$-attractors: phenomenology and observational
constraints}

\author[a, b]{Carlos García-García,}
\emailAdd{carlosgarcia@iff.csic.es}
\author[c, d]{Eric V. Linder,}
\emailAdd{evlinder@lbl.gov}
\author[a, b]{Pilar Ruíz-Lapuente}
\emailAdd{pilar@icc.ub.edu}
\author[c, e]{and Miguel Zumalacárregui}
\emailAdd{miguelzuma@berkeley.edu}

\affiliation[a]{
  Instituto de Física Fundamental, Consejo Superior de Investigaciones
Científicas, c/. Serrano 121, E–28006, Madrid, Spain }
\affiliation[b]{
  Institut de Ci\`{e}ncies del Cosmos (UB–IEEC), c/. Martí i Franqués 1,
E–08028, Barcelona, Spain}
\affiliation[c]{
Berkeley Center for Cosmological Physics and Berkeley Lab, University of California, Berkeley, CA 94720, USA
}
\affiliation[d]{
  Energetic Cosmos Laboratory, Nazarbayev University, Astana, Kazakhstan 010000
}
\affiliation[e]{
Institut de Physique Th\'eorique, Universit\'e  Paris Saclay CEA, CNRS, 91191
Gif-sur-Yvette, France
}

\abstract{The possibility of linking inflation and late cosmic accelerated
  expansion using the $\alpha$-attractor models has received increasing
  attention due to their physical motivation. In the early universe,
  $\alpha$-attractors provide an inflationary mechanism compatible with Planck
  satellite CMB observations and predictive for future gravitational wave CMB
  modes.  Additionally $\alpha$-attractors can be written as quintessence
  models with a potential that connects a power law regime with a plateau or
  uplifted exponential, allowing a late cosmic accelerated expansion that can
  mimic behavior near a cosmological constant.  In this paper we study a
  generalized dark energy $\alpha$-attractor model.  We thoroughly investigate
  its phenomenology, including the role of all model parameters and the
  possibility of large-scale tachyonic instability clustering.  We verify the
  relation that $1+w\sim 1/\alpha$ (while the gravitational wave power
  $r\sim\alpha$) so these models predict that a signature should appear in
  either the primordial B-modes or in late time deviation from a cosmological
  constant.  We constrain the model parameters with current datasets,
  including the cosmic microwave background (Planck 2015 angular power
  spectrum, polarization and lensing), baryon acoustic oscillations (BOSS
  DR12) and supernovae (Pantheon compressed).  Our results show that expansion
  histories close to a cosmological constant exist in large regions of the
parameter space, not requiring a fine-tuning of the parameters or initial
conditions.}

\begin{document}
\maketitle

\section{Introduction}
\label{S: Introduction}

In the last few years, different inflationary models were realized to yield
similar values for the primordial scalar perturbation ratio, $n_s$, and the
gravitational wave tensor to scalar ratio, $r$, for a wide range of inflation
potentials. These models predict that for $N$ e-folds of inflation, up to
leading order \cite{Galante:2014ifa}, 
\begin{equation}
  n_s =  1 - 2 N^{-1} \quad \mbox{ and } \quad r = 12 \alpha N^{-2}\,,
\end{equation}
being compatible with WMAP \cite{Hinshaw:2012aka} and Planck 
\cite{Ade:2015lrj} cosmic microwave background (CMB) observations. The reason why all of them have similar
predictions is a second order pole in the kinetic coefficient of the Einstein frame
Lagrangian; i.e. with no coupling to the Ricci scalar and a non-canonical
kinetic coefficient for the scalar \cite{Galante:2014ifa}:
\begin{equation}
  {\mathcal L}=\sqrt{-g}\,\left[\frac{1}{2}M_P^2 R-\frac{\alpha}{(1-\varphi^2/6)^2}\,\frac{1}{2} (\partial\varphi)^2
  -\alpha f^2\left(\frac{\varphi}{\sqrt{6}}\right)\right]\ ,
  \label{eq:L-original}
\end{equation}
where $M_P$ is the Planck mass, $\alpha$ is a parameter and $\alpha f^2$ is
the potential function dependent on the field $\varphi$ which is measured in
$M_P$ units. This Lagrangian is obtained from an inflationary multifield
Lagrangian with a locally conformal transformation symmetry, once the extra
degree of freedom associated to this symmetry is gauge fixed and the potential
function $f$ is required to be real \cite{Kallosh:2013hoa}. 

The field redefinition $\phi=\sqrt{6\alpha}\,\arctanh(\varphi/\sqrt{6})$
makes the scalar field's kinetic coefficient canonical, allowing to write the theory
as a quintessence model
\begin{equation}
{\mathcal L}=\sqrt{-g}\,\left[\frac{1}{2}M_P^2 R- \frac{1}{2}
(\partial\phi)^2 -\alpha f^2(x)\right]\ ,
\label{eq:L}
\end{equation}
where $x = \tanh(\phi/\sqrt{6\alpha})$. Now, the field space is expanded in
the connected region of $\varphi$ since this transformation pushes the limits
of the original field, $\varphi \in (-\sqrt{6}, \sqrt{6})$, towards $\pm
\infty$ in the transformed field, $\phi \in (-\infty, \infty)$.

In the dark energy context, $\alpha$-attractors have been gaining attention
due to their possibility for linking both inflationary
and present accelerated expansions. In the same way as their predictions in
inflation are in good agreement with the latest cosmological observations, as
quintessence models they can also produce a late accelerated expansion
compatible with present measurements. 
Their potential connects power law scalar field potentials near the 
minimum $x=0$ to a 
cosmological constant, in the form of a plateau as $x\to\infty$.  A series of works based on
this idea have appeared recently.  The connection of both cosmological epochs
is studied in Ref.~\cite{Dimopoulos:2017zvq}, where a
particular potential with two plateau regimes allows both inflation and a near
cosmological constant expansion. This work was extended in 
Ref.~\cite{Dimopoulos:2017tud} by computing the reheating era and setting
constraints on the model parameter space (see also \cite{Akrami:2017cir, Casas:2017wjh}).

Concentrating on late time acceleration and the $\alpha$-attractors as 
dark energy, Ref.~\cite{Linder:2015qxa} related these models to the thawing 
and freezing classes of dark energy and generalized the two most common 
forms of the potential to a unified form, which we will use here. Other 
works have also examined the dark energy applications, e.g.\ 
\cite{Shahalam:2016juu,Bag:2017vjp}. 

Other investigations include the use of $\alpha$-attractors for 
dark matter \cite{Mishra:2017ehw}, and the relation of $\alpha$-attractors 
with $f(R)$ gravity, extending the original connection 
of $\alpha$-attractors with Starobinsky $R^2$ gravity
\cite{Starobinsky:1980te}, e.g.\ \cite{Odintsov:2016vzz, Miranda:2017juz}.

We focus here on testing the $\alpha$-attractor dark energy model against 
different cosmological observations and physical understanding of the 
constraints on the parameter space. We use the generalized potential of 
Ref.~\cite{Linder:2015qxa} and allow the initial field value to vary, 
since fixing it restricts the phenomenology and can bias the results. 
Furthermore, we will thoroughly investigate the 
dependence of the model on each parameter and its initial value conditions
showing that cosmological constant-like solutions are generic and do not
require any fine-tuning of the model parameters and initial conditions. In
addition, we will investigate the tachyonic instabilities noted in
Ref.~\cite{Linder:2015qxa} to see if there is any signatures of interesting
phenomenology, such as clustering dark energy, and if they can cause an
observable imprint. 

In Section \ref{S:Model}, we briefly 
review the model proposed in Ref.~\cite{Linder:2015qxa} and carefully
examine the theory dependence on each parameter (Sections \ref{S:alpha} and
\ref{S:pn}). We investigate in Section \ref{S:obs} how the observables 
change with the model parameters, including a quantitative assessment of 
the tachyonic instability phenomenon in terms of an observable signature. 
In Section~\ref{S:MCMC} the model is confronted against observational data 
from CMB Planck 2015 \cite{Ade:2015rim}, baryon acoustic oscillation (BAO) 
DR12 \cite{Alam:2016hwk}, and the supernove Type Ia distances in terms of 
binned $E(z) = H(z)/H_0$ \cite{Riess:2017lxs}. Section~\ref{S:data} describes 
these datasets and why we chose them, while in section \ref{S:priors} we 
present the priors to be used in the Bayesian study of 
the next two subsections. In section \ref{S:MCMC1}, we compare the
Starobinsky form of the potential with $\Lambda$CDM, leaving just one 
parameter (the scaling $\alpha$ itself) free in order to assess if this 
form, and the Starobinsky value $\alpha=1$ is favored. In 
Section~\ref{S:MCMC2} we then free every parameter to obtain the full
$\alpha$-attractor generalized model posterior distribution. We conclude 
in Section~\ref{S:conclusion}.

\section{The model}
\label{S:Model}

\begin{figure}[t]
  \centering
  \includegraphics[width=0.4\textwidth]{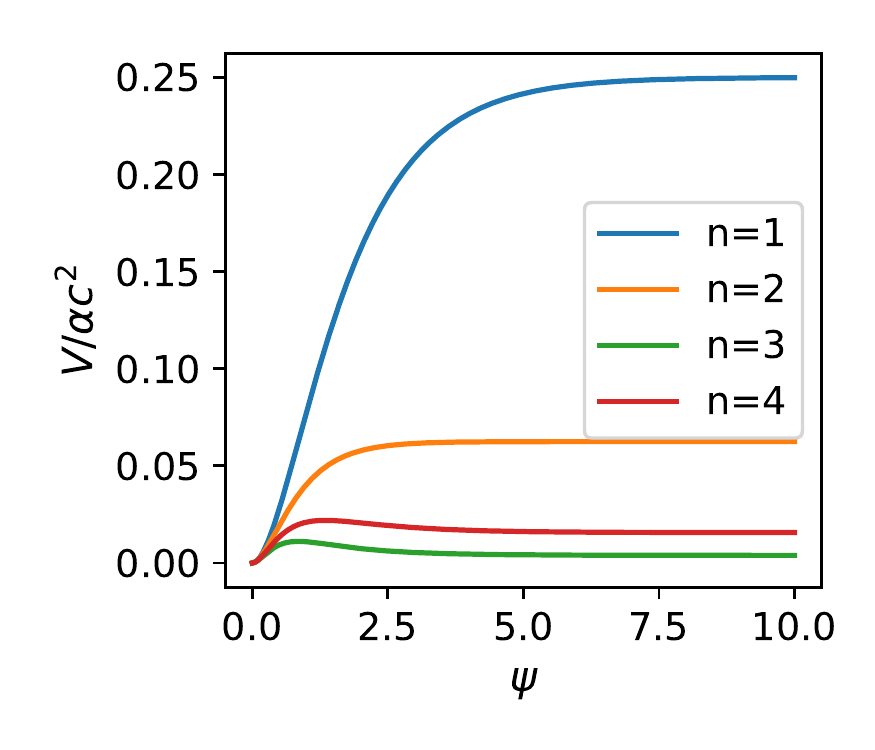}
  ~
  \includegraphics[width=0.4\textwidth]{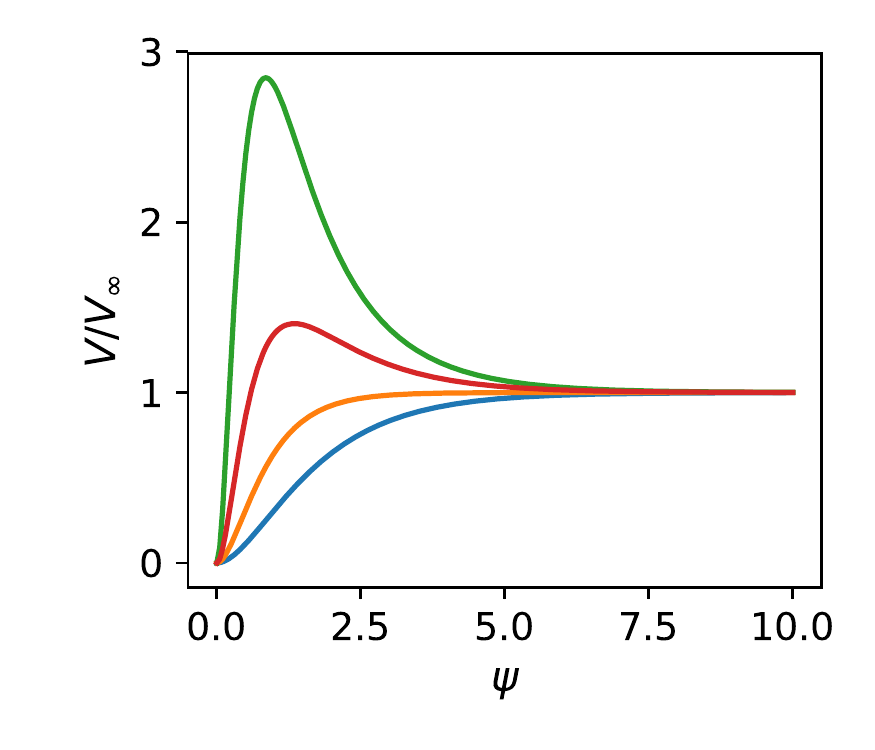}
  \caption{Generalized $\alpha$-attractor potential for different values of
  $n$, fixing $p=2$. For $n>2$, there is a maximum. The maximum strengthens 
with $n$ if the potential is normalized to its asymptotic amplitude.} 
  \label{fig:V}
\end{figure}

We will work on the generalized $\alpha$-attractor potential from
\cite{Linder:2015qxa}. In the language of equation \ref{eq:L}, $V(\phi) =
\alpha f^2(x)$ is the field potential and, in this case, is given by
\begin{equation}
  V(x)=\al c^2\,\frac{x^p}{(1+x)^{2n}} = \al c^2 2^{-2n}(1-y)^p (1+y)^{2n-p} \ , 
  \label{eq:V}
\end{equation}
with $c$, $p$, $n$ constant parameters and $y \equiv e^{-2\phi/\sqrt{6
\alpha}}$. The case $\al = 1$, $n = 1$, $p = 2$ corresponds to the Starobinsky
model \cite{Whitt:1984pd, Maeda:1987xf, Barrow:1988xi}, working in natural units,
i.e.\ reduced Planck mass $M_P = 1$ and speed of light, $c = 1$. 
We will work on a flat geometry motivated by inflation. 
Figure~\ref{fig:V} shows the potential for different values of
$n$ using the variable $\psi \equiv \phi/\sqrt{\alpha}$, the scaled scalar
field, since it is actually what determines the value of $V(x)$. 
Note how $n$ controls the transition from the flat plateau to the 
monomial-shaped minimum.

Let us briefly summarize the potential properties before studying thoroughly
the dependence on each parameter. The potential interpolates from a power law
potential with index $p$ to a cosmological constant fixed value, basically 
an uplifted (negative) exponential potential, for positive values of the 
field: 
\begin{eqnarray}
  V(|\psi|\ll \sqrt{6}) &\approx& \alpha c^2 6^{-p/2}\,\psi^p \ ,
  \label{eq:Vflow}\\
  V(\psi \gg \sqrt{6}) &\approx& \al c^2\,2^{-2n}\,
  \left[1-2(p-n)\,e^{-2\psi/\sqrt{6}}\right] \xrightarrow[\psi \to \infty]{}
  \frac{\alpha c^2}{2^{2n}} \ .
  \label{eq:Vfgtr}
\end{eqnarray}
We see that the amount of dark energy in the Universe will be determined
by the potential amplitude, $\alpha c^2$, whose characteristic scale will
depend on $n$ as ultimate responsible of the height of the plateau --- for a
model with $p = 2\,, \alpha=1$ and $n = 0,\, 3,\, 5$, the amplitude $\alpha
c^2 \sim \unit[10^{-7},\, 10^{-6},\, 10^{-5}]{Mpc^{-2}}$, respectively. Note
there is no true cosmological constant: the potential is zero at the minimum. 

The $\alpha$-attractor potential has a maximum at $\xmax = p/(2n - p)$ given
$n > p$ \cite{Linder:2015qxa}. Fields starting (from rest) at $x>\xmax$ 
will roll towards infinity and the asymptotic constant potential, 
i.e.\ a de Sitter solution, while asymptotically
freezing. On the other hand, for $x<\xmax$ 
the field would roll toward the origin along the plateau and eventually 
(possibly in the future) down toward the zero minimum. However, if it 
rolls too far off the plateau the kinetic energy rises, forcing the 
equation of state $w$ far from $-1$ and it would not
be a viable dark energy model today. Therefore we are not concerned with 
fields rolling past the minimum and so only need to deal with positive 
field values. Fields with $x<\xmax$ basically act like 
thawing models; they depart from a cosmological constant like behavior. 

So far we have talked about the known properties of this model, already
studied in Ref. \cite{Linder:2015qxa}.  Now, we will start our detailed study on
its dependence on each parameter. We will vary $\alpha$, $n$ and $p$. 
The parameter $c$ is fixed by the closure relation $1 = \sum_i \Omega_i$, 
where $\Omega_i$ is the fractional energy density, i.e.\ the ratio of the 
energy density of component $i$ (e.g.\ matter, dark energy, radiation) 
to the critical density. To solve the field evolution 
\begin{equation}
  \ddot \phi + 3 H \dot\phi + V_{\phi} = 0\ ,
  \label{eq:field-eq}
\end{equation} 
where $V_\phi=dV/d\phi$, 
we also need to specify the initial value of the field $\phi_{ini}$ and 
its velocity $\dot \phi_{ini}$. For a field starting on the nearly flat 
plateau (and if it starts off the plateau it is not using the 
$\alpha$-attractor characteristics) 
the Hubble friction will freeze the field at early times (we start at 
$z_{ini}=10^{14}$ in the radiation era) and so we take $\dot \phi_{ini} = 0$. 
Thus $\phi_{ini}$ is the only further parameter. 
To solve the coupled differential equations that govern the cosmological 
history we use the hi\_class Boltzmann
code\footnote{\url{http://hiclass-code.net}}
\cite{Blas:2011rf, Zumalacarregui:2016pph}.

\subsection{Dependence on the scaling of the potential ($\alpha$)}
\label{S:alpha}

We start our study on the effect of the different parameters varying the
simplest one: the scaling $\alpha$. This has close connections with the 
underlying particle physics (e.g.\ supergravity or conformal field theory 
model). Note that $\alpha$ scales the field value and the potential 
amplitude, but cannot be removed by a field redefinition since it does 
not appear in the kinetic terms $\dot\phi^2/2$. 

\begin{figure}[t]
  \centering
  \includegraphics[width=\textwidth]{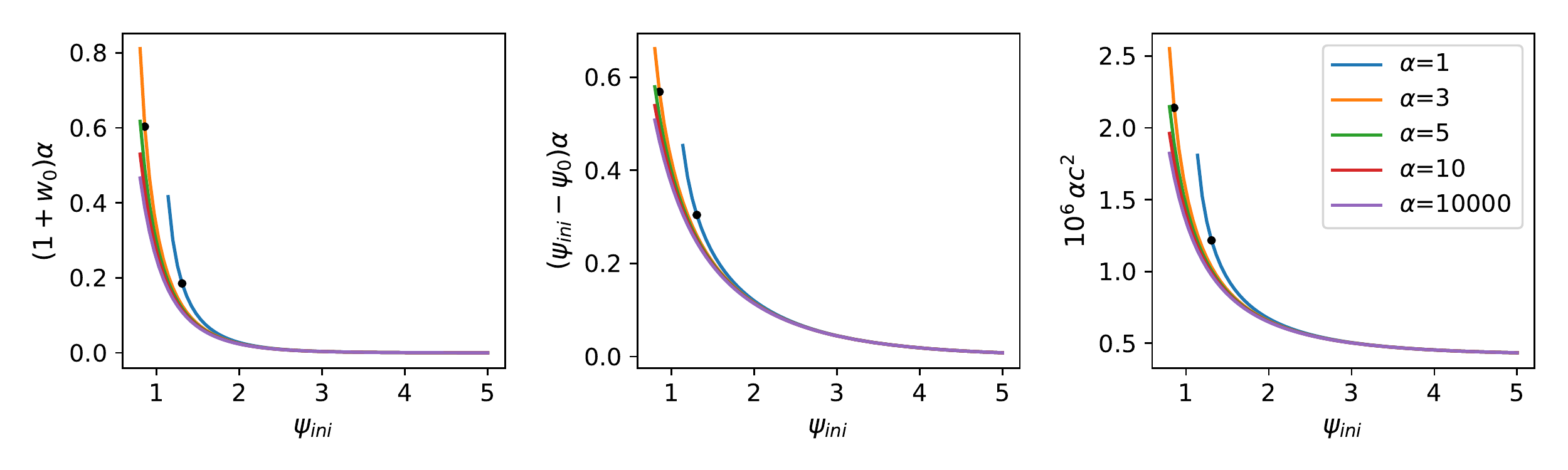}
  \caption{The dynamical quantities $(1+w_0)\alpha$ and 
$(\psi_{ini} - \psi_0)\alpha$ are nearly independent of $\alpha$ for values 
of $\psi_{ini}$ that deliver $1+w_0<-0.8$ (marked by black dots). 
Other parameters are fixed to $p=2$ and $n=1$, with  $c$ fixed by
  the density closure relation. } 
  \label{fig:continuous-fini_alpha-prop}
\end{figure}

Generally, larger values of $\alpha$ bring the model closer to $\Lambda$CDM, 
as the potential dominates more over the kinetic energy and the plateau 
is stretched out longer for a given value of $\psi$. We can make this more 
quantitative by employing the flow formalism 
for a thawing field \cite{Cahn:2008gk}, 
\begin{equation}
  1 + w = \frac{4}{27} \left(\frac{V_\phi}{V}\right)^2 \Omega_{DE} +
  \mathcal{O}(V_{,\phi\phi})
  \label{eq:w-prop-alpha}
\end{equation}
Since for fields on the plateau, $V_\phi/V\sim 1/\sqrt{\alpha}$, then 
$1+w \sim 1/\alpha$. We have numerically
checked this relation holds until quite recent times and plotted
$(1+w_0)\alpha$ in Figure~\ref{fig:continuous-fini_alpha-prop}. Note that in
the quintessential inflation model of Ref.~\cite{Akrami:2017cir} they also find
$1+w\sim 1/\alpha$, while the CMB tensor to scalar ratio $r\sim\alpha$ so a
physics signature is present either in $r$ (if $\alpha$ is large) or $w$ (if
$\alpha$ is small).  In addition, we can see that for thawing models the field
evolution $(\psi_{ini} - \psi_0)\sim 1/\alpha$, where $\psi_0$ is the value
today.  This relation comes from 
\begin{equation}
  (\phi_0 - \phi_{ini}) = \int_{ini}^0 dt\, \dot\phi \approx 
\int_{ini}^0 dt\, \sqrt{\rho_{DE}(1+w)}\, .
  \label{eq:DeltaPhi-prop-alpha}
\end{equation}
Since the dark energy density changes little for thawing models, one can
see that $\sqrt{\alpha}(\psi_{ini} - \psi_0) \sim \sqrt{1+w_0}$ \cite{Linder:2015zxa}.
Effectively, Figure~\ref{fig:continuous-fini_alpha-prop} shows that for viable
models (those with $w_0<-0.8$) these relations hold quite well.

Note the initial value of the field plays an important role, with fields
with high initial values mimicking the results of a cosmological constant. 
That is, $w_0\approx-1$ and $\psi_{ini}-\psi_0\approx0$. 
This is seen in Figure~\ref{fig:continuous-fini_alpha-prop}. 
Note that when $p>n$ (here $p=2$, $n=1$), 
the potential does not have a maximum and the field will
always roll down towards $\psi=0$. However, sufficiently large initial 
fields stay on the plateau nearly frozen for the whole evolution
history of the Universe. For initial values of the field closer to $\psi=0$,
the parameter $\alpha$ determines when the field starts rolling down and, as a
consequence, how fast it moves and how far the equation of state parameter is
from the cosmological constant solution. Larger $\alpha$'s slow
the field down, keeping it closer to cosmological constant behavior. 

The case with $p < n$ has a maximum that causes 
a different phenomenology depending on the relative size of
$\psi_{ini}$ and $\psi_{max}$ so that any field starting at high values 
($\psi_{ini} > \psi_{max}$) would roll towards the de Sitter attractor.  
This case will be studied in detail in Section~\ref{S:pn}
and is shown in Figure~\ref{fig:continuous-fini_n.png}. 

Finally, the potential amplitude $\alpha c^2$, 
responsible for the amount of dark energy in the
Universe, which is kept fixed, needs to be adjusted to compensate the loss
that comes from the evolving part of the potential, $V/\alpha c^2$. 
This is shown in the rightmost panel of Figure~\ref{fig:continuous-fini_alpha-prop}. 
It is fairly insensitive to $\psi_{ini}$, especially for large $\psi_{ini}$ 
where the field stays nearly frozen. As Eq.~(\ref{eq:Vfgtr}) shows, once 
the amplitude is accounted for, the remaining form of the potential is 
insensitive to $\alpha$, so the curves all lie together. 

Thus, models will be close to $\Lambda$CDM if they either have large 
$\alpha$ (recall $1+w_0\sim 1/\alpha$) or large $\psi_{ini}$. Since 
allowing them to get larger and larger will give the same physical 
results, in Section~\ref{S:priors} we pay careful attention to priors 
for the Monte Carlo analysis.

\subsection{Dependence on the shape of the potential ($p$ and $n$)}
\label{S:pn}

In this section we will study how the exponents, $p$ and $n$, change the field
evolution history. We treat them together because, leaving aside the low and 
high $\psi$ regimes, where the potential is governed by, respectively, 
$p$ and $n$ separately (Equations~\ref{eq:Vfgtr} and \ref{eq:Vflow}), 
in the most interesting, intermediate $\psi$ values it is their relative 
size that matters most. 

\begin{figure}[t]
  \centering
  \includegraphics[width=\textwidth]{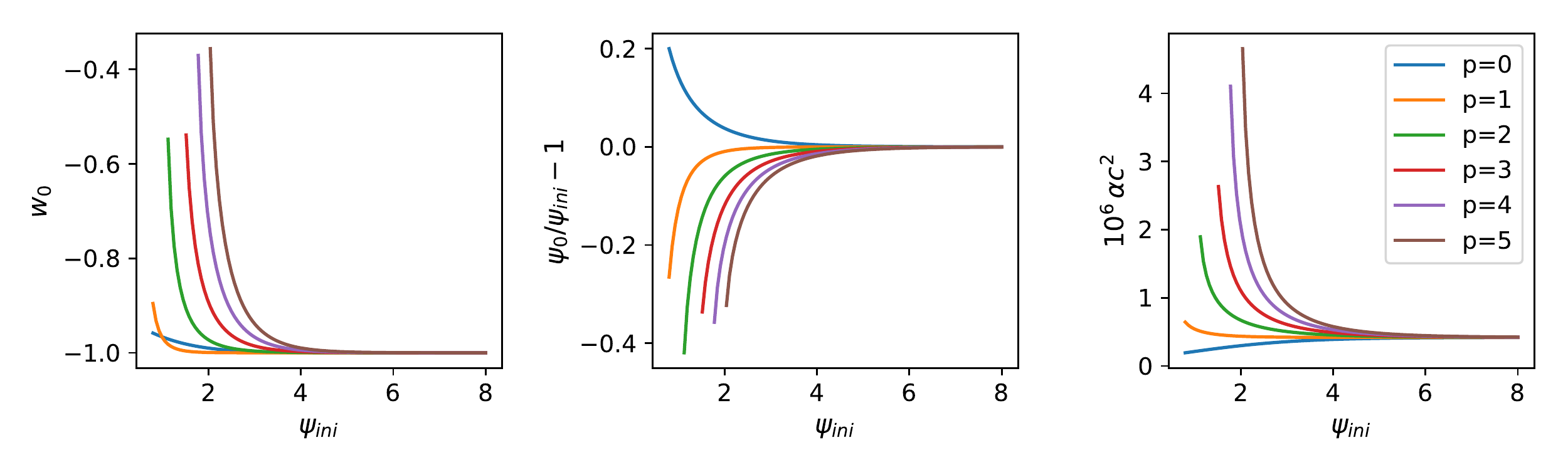}
  \caption{
The values of $w_0$, $\psi_0/\psi_{ini} - 1$ and $\alpha c^2$ are plotted 
vs $\psi_{ini}$ for various values of $p$, for a fiducial model with 
$\alpha=1$ and $n=1$. Except 
    for $p=0$, for which the potential (equation \ref{eq:V}) is monotonically
    decreasing, the 
    exponent $p$ determines the low-$\psi$ shape of the potential. For 
 $p\geq n$, the steepness grows with $p$, making the
    field evolve faster (and earlier). The potential amplitude $\alpha c^2$ 
varies to compensate for the field evolution in order to preserve 
the same present dark energy density. 
Note that viable models with $w_0<-0.8$ need small $p$ or
    high $\psi_{ini}$.} 
  \label{fig:continuous-fini_p.png}
\end{figure}

For completeness, recall that $p$ governs the low $\psi$ regime (equation
\ref{eq:Vflow}) as $V \propto \psi^p$, and the slope of the
potential (for a given $n$) in the transition between the plateau and 
the power law regime. The
field evolution will be faster and earlier 
with larger $p$. Nevertheless, this prescription is only
valid in the low-$\psi$ regime and in the case $p>n$. The case 
with $n>p$, which will
be analyzed later, is different because of the appearance of a maximum, whose 
size, position and steepness is determined by their relative size. The only
different case is $p=0$ because it is the only configuration that exclusively
allows the field to grow towards the de Sitter attractor, since the 
potential is monotonically 
decreasing. This analysis is confirmed by numerical solution of the evolution
equations, as seen for $\alpha=1$
and $n=1$, in Figure~\ref{fig:continuous-fini_p.png}. 

\begin{figure}[t]
  \centering
  \includegraphics[width=\textwidth]{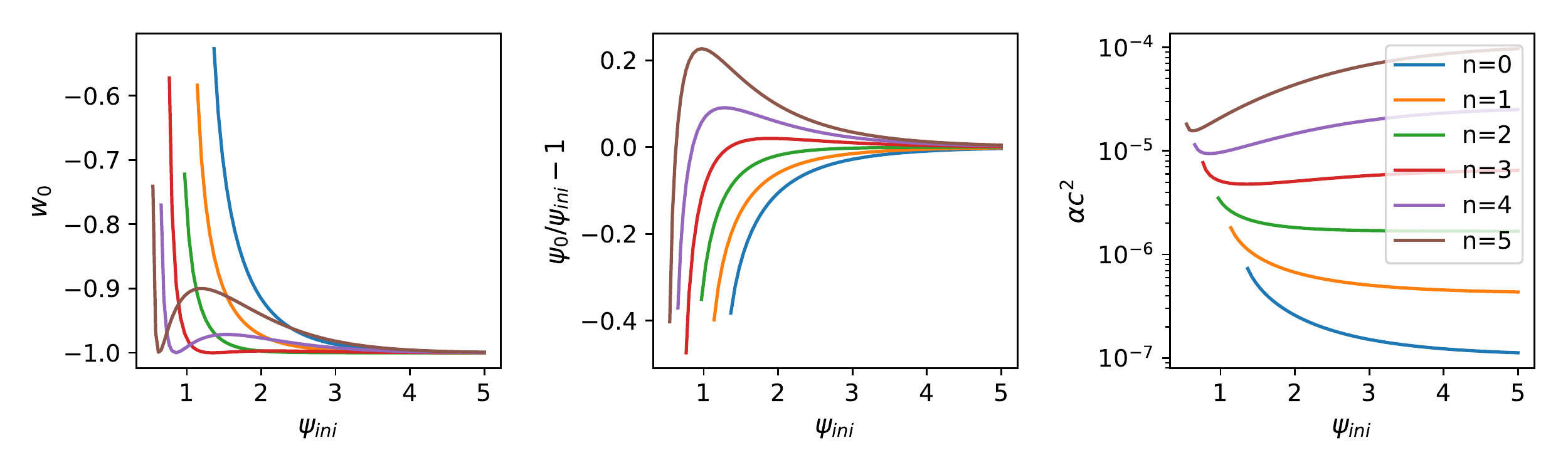}
  \caption{
As Figure~\ref{fig:continuous-fini_p.png} but varying with respect to $n$,
with a fiducial model $\alpha=1$ and $p=2$.  All curves end at the lowest
value that gives $\Omega_{DE} \sim 0.7$ today. The relative size of $n$
respect to $p$ changes the potential slope on the plateau: the closer they
are, the less pronounced the slope is. This means that as $n$ goes to $p=2$,
the field rolls down later and slower.  The potential amplitude $\alpha c^2$
varies to compensate for the amplitude reduction $2^{-2n}$ in
Equation~\ref{eq:Vfgtr} in order to preserve the same present dark energy
density.  Note that viable models with $w_0<-0.8$ need $n\approx p$ or high
$\psi_{ini}$.} 
  \label{fig:continuous-fini_n.png}
\end{figure}

The exponent $n$, instead, fixes the asymptotic behavior of $V(\psi)$ as $V
\rightarrow \alpha c^2 2^{-2n}$ (equation \ref{eq:Vfgtr}) having a direct
impact on the potential amplitude. This means that for a fixed
amount of present dark energy density, the value of $\alpha c^2$ 
must be modified in order
to correct the  deficit caused by $n$; thus it shifts up or down 
as seen in Figure~\ref{fig:continuous-fini_n.png}. 

We can also use figure \ref{fig:continuous-fini_n.png} to study the
interdependence of both exponents in the intermediate $\psi$ values since
$p=2$ and $n$ ranges from 0 to 5 so that there are solutions with ($p<n$)
and without ($p>n$) a potential maximum. In the later case, the potential
plateau is slightly inclined towards $V=0$ and the transition regime that
connects it with the power law regime is steeper as $p-n$ grows. This
expression  also governs the size, slope and position of the maximum when
$n>p$. On the one hand, its position is given by $x_{max} = p/(2n - p)$, so
that increasing $n$ shifts the potential maximum toward $\psi=0$.
Quantitatively, for the studied cases with $n=3,\,4,\,5$, the maximum is
located at $\psi_{max} = 1.35,\, 0.85,\, 0.63$ . On the other hand, fixing the
dark energy content makes the peak higher and steeper: 
for $n=3,\,4,\,5$ the ratio of the potential maximum to its asymptotic value 
is $V_{max}/V_{\infty} = 1.40, 2.84, 6.87$, respectively.

Thus, fields starting at $\psi_{ini} > \psi_{max}$ roll down toward
$\infty$ with a velocity dependent on its proximity to the maximum. While
fields with $\psi_{ini}\gg\psi_{max}$ remains almost frozen ($w_0 \sim -1$) on
the plateau, those with $\psi_{ini} \sim \psi_{*}$ ($\psi_{*}$ is the
inflection point at $\psi > \psi_{max}$), where the slope is maximal, speed
up, departing from the cosmological constant solution ($w_{0}$ grows). For 
$\psi_{ini}$ closer to $\psi_{max}$ 
the field feels a weaker force and its evolution is slower ($w_0$ decreases),
having as a limit case, $\psi_{ini} = \psi_{max}$, where the field remains 
frozen for the whole evolution history of the Universe ($w = -1$). Starting at
$\psi_{ini} < \psi_{max}$ the field rolls down again, but this time toward
$\psi = 0$. The closer to the minimum it starts, the faster it evolves ($w_0$
grows quickly). Note that the shift in the 
maximum reduces the
available space at $\psi_{ini} < \psi_{max}$ as $n$ grows, e.g.\ giving 
the shift of the minimum to the left in 
Figure~\ref{fig:continuous-fini_n.png}.

\subsection{Model Predictions and Observables}
\label{S:obs}

Now we will focus on the phenomenological predictions of the model given by
equation~\ref{eq:V}. Besides the effects on the background expansion, in 
Ref.~\cite{Linder:2015qxa} it was suggested that, as a
consequence of having $m^2 = V'' < 0$ near the edge of the plateau, 
one might find some interesting phenomenology, such as clustering dark 
energy. The field perturbations $\delta\phi_k$ in momentum space 
becomes \cite{Peter:2013avv}
\begin{equation}
  \delta\ddot\phi_k + 3H\delta\dot\phi_k + \left( \frac{k^2}{a^2} - m^2
  \right)\delta\phi_k = 4\dot\phi_k \dot\Phi - 2V_{,\phi}\Phi\, ,
  \label{eq:pert-eq}
\end{equation} 
in standard Newtonian gauge notation, where the metric perturbations 
$\Phi = \Psi$. 

Since perturbations start growing significantly at horizon entering, we will
just consider $k>H$ modes. In addition, equation \ref{eq:pert-eq} tell us that
the mass term must be $|m^2| \gtrsim k^2/a^2$ in order to change the
perturbation growth, bounding the scales sensitive to the imaginary mass to 
\begin{equation} 
  \frac{|m^2|}{H^2} \gtrsim \frac{k^2}{a^2 H^2} > 1\,.
  \label{eq:k_mass} 
\end{equation}
Given that the field starts evolving at late times ($z \sim O(1)-O(10)$) and
that the field's mass will not be extremely high, the affected wavenumber will
be $k\approx\unit[10^{-3}]{Mpc^{-1}}$, i.e.\ much larger scales than where 
precision clustering data lies.  In addition, late time evolution implies 
that perturbations will not have much time to grow. 

\begin{figure}[t]
  \centering
  \includegraphics[width=\textwidth]{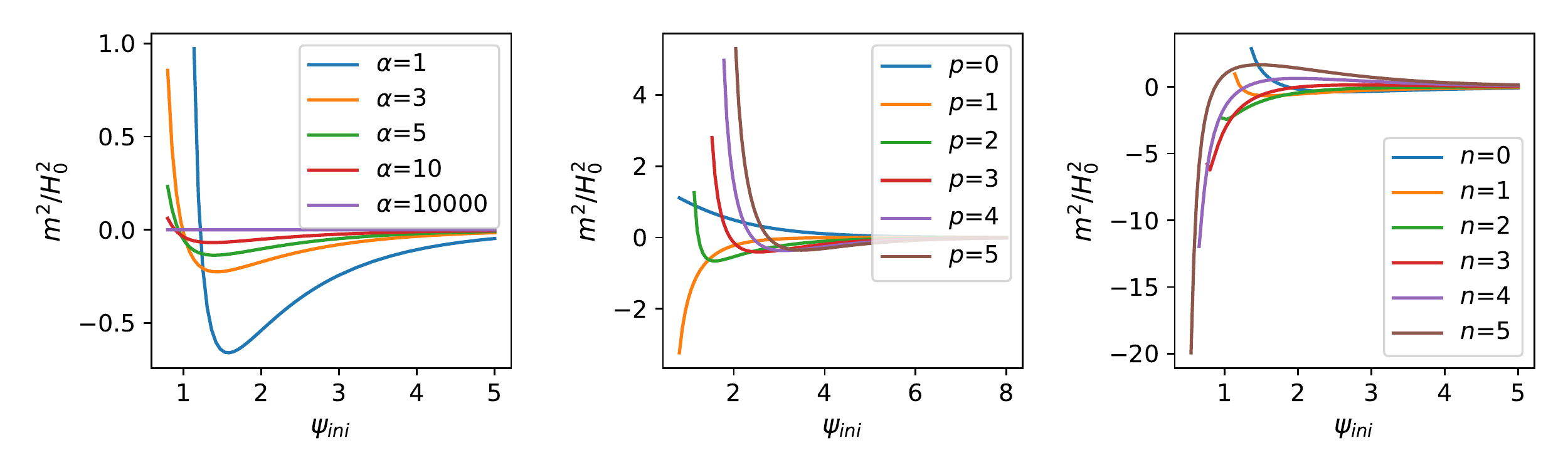}
  \caption{
The mass squared $m^2/H_0^2$ can go negative, as is standard for hilltop 
models, but general not deeply (relative to $H^2$) or for long. 
When $n<p$, there are two regions, delimited by the inflection point in
the transition between the power law regime and the plateau, where the mass
squared is positive or negative, respectively. On the contrary, if $n>p$, the
appearance of a maximum means that $m^2<0$ is bounded around it by the two
inflection points. As commented in section~\ref{S:pn}, $n$ has a great 
impact on the relative size of the maximum
respect to the plateau, making it also the parameter with greatest
impact on $m^2$.}
\label{fig:continuous-fini_m2H2.png}
\end{figure}

As a consequence, observing some effect requires a sufficiently
large negative mass squared. We have computed the present value of the mass
squared term for each model studied in the previous section, and plotted it in
figure \ref{fig:continuous-fini_m2H2.png}. One can read from it that models
with exponents $n<p$ have only two regions, separated by the inflection point
in the transition zone, so that $m^2<0$ is expected for high $\psi_{ini}$ and
$m^2>0$ at low $\psi_{ini}$. On the contrary, as $n>p$ implies the appearance
of a maximum, the region with $m^2<0$ is bounded by the inflection points. It
is important to remark that, given that $n$ makes the maximum steeper as it
grows, high $n$ can give rise to 
sufficiently negative $m^2$, which, in theory, could
be potentially noticeable.  Nevertheless, for the sensible models studied
here, with no extreme exponents, and compatible with the dark energy density
observations, which in the end fixes the potential amplitude, we have found
that all models have a similar perturbation growth, just varying a little bit
close to the present. 

\begin{figure}[t]
  \centering
  \includegraphics[width=0.33\textwidth]{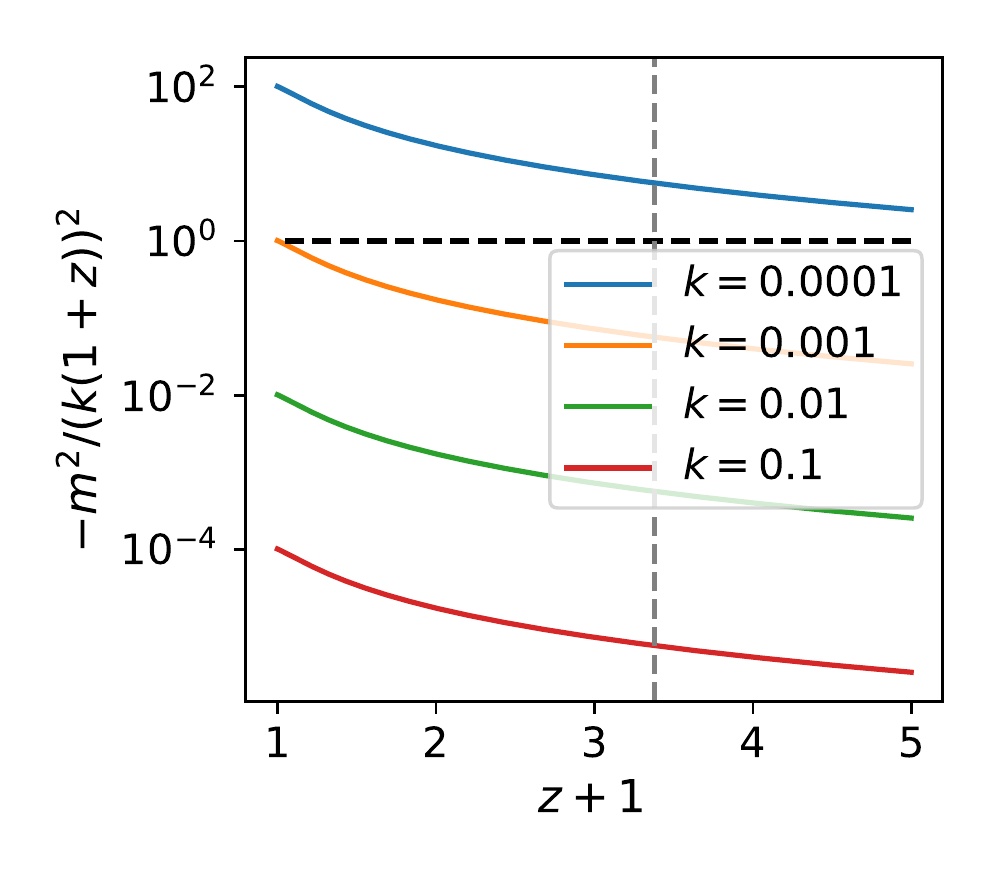}\hfill
  \includegraphics[width=0.33\textwidth]{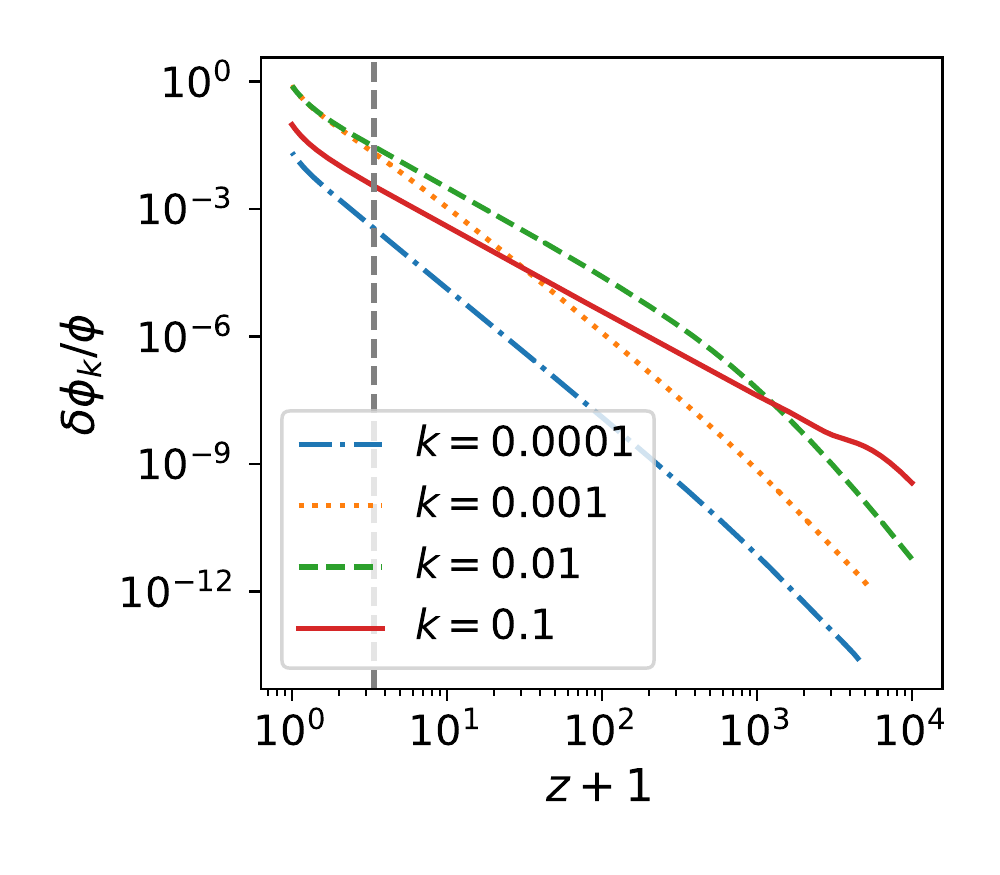}\hfill
  \includegraphics[width=0.33\textwidth]{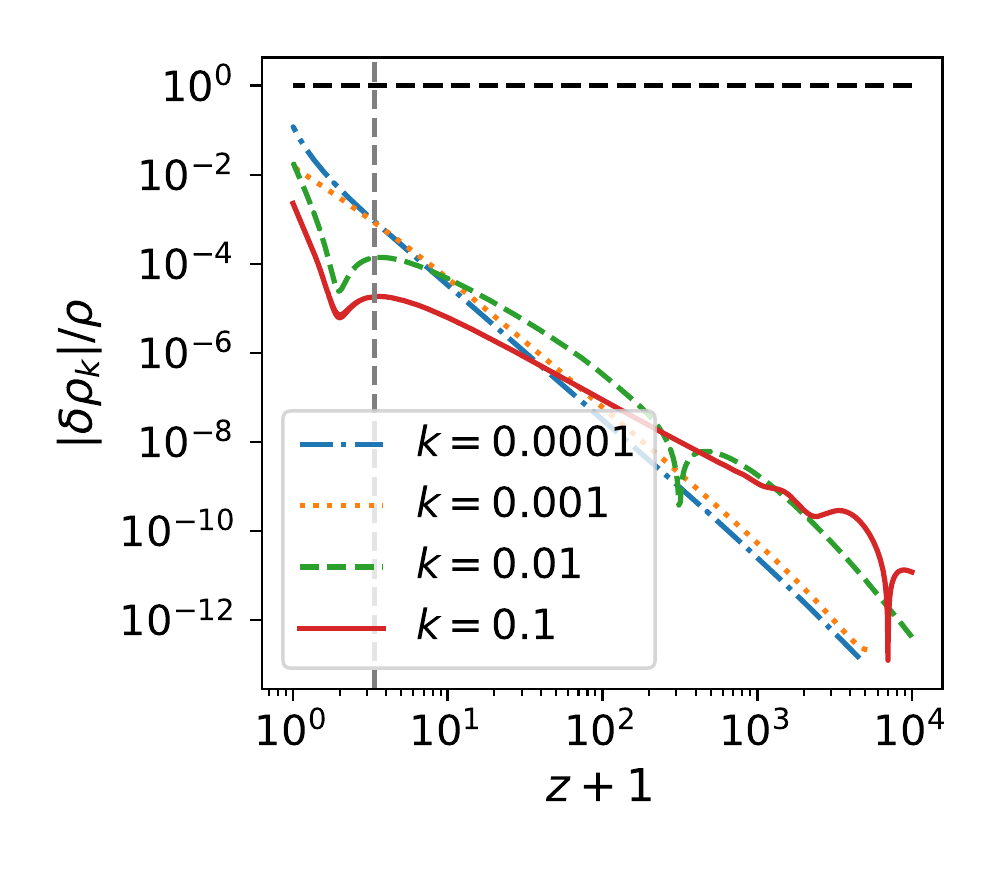}
  \caption{Redshift evolution is shown of the mass squared relative 
to the $k^2/a^2$ term
    (left), the normalized field perturbation (middle), and the 
normalized density
    perturbation for the model $\alpha=1$, $\psi_{ini} = 0.542$, $p = 2$ and
    $n = 5$, the extreme model with $m_0^2/H_0^2 \approx -20$. 
The vertical dashed line marks the redshift
  when $|m^2|/H^2 = 1$ and the horizontal one shows $|\delta\rho_k/\rho| = 1$.
    $k$ is in units of $\unit[]{Mpc^{-1}}$. Recall that the mode with
    $k=\unit[10^{-4}]{Mpc^{-1}}$ has not yet entered in the horizon. Dark
    energy density perturbation modes are negligible with respect to its background
  value during almost the whole evolution history even for this extreme model.}
  \label{fig:m2-pert}
\end{figure}

Figure \ref{fig:m2-pert} illustrates that even the extreme model with 
with $m^2/H_0^2 \sim -20$ (that indeed has unviable equation of state 
$w(a)\approx -0.74-1.9(1-a)$) shows that perturbations have little effect 
on the dark energy density or field value for the cosmic history up to 
the present.

\begin{figure}[t]
  \centering
  \includegraphics[width=0.333\textwidth]{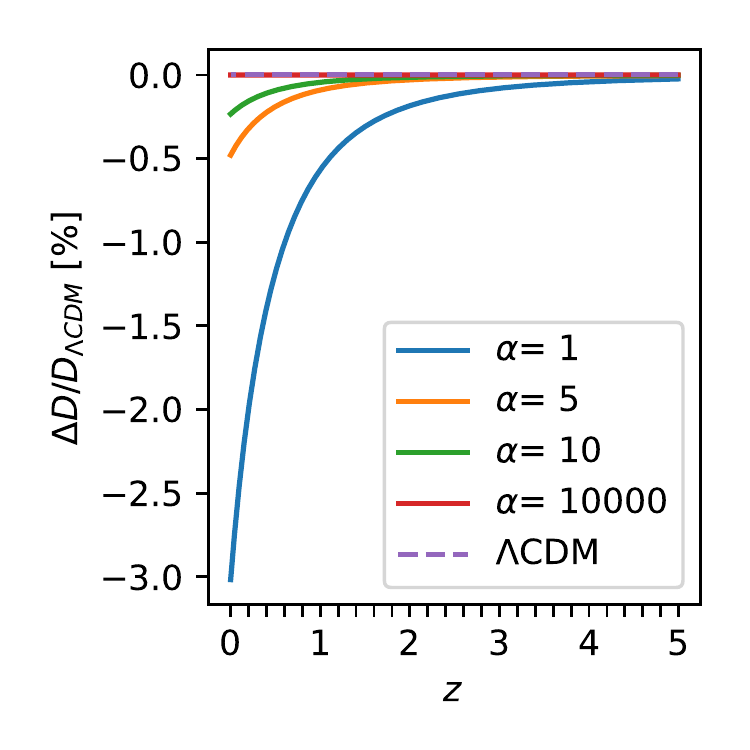}\hfill
  \includegraphics[width=0.333\textwidth]{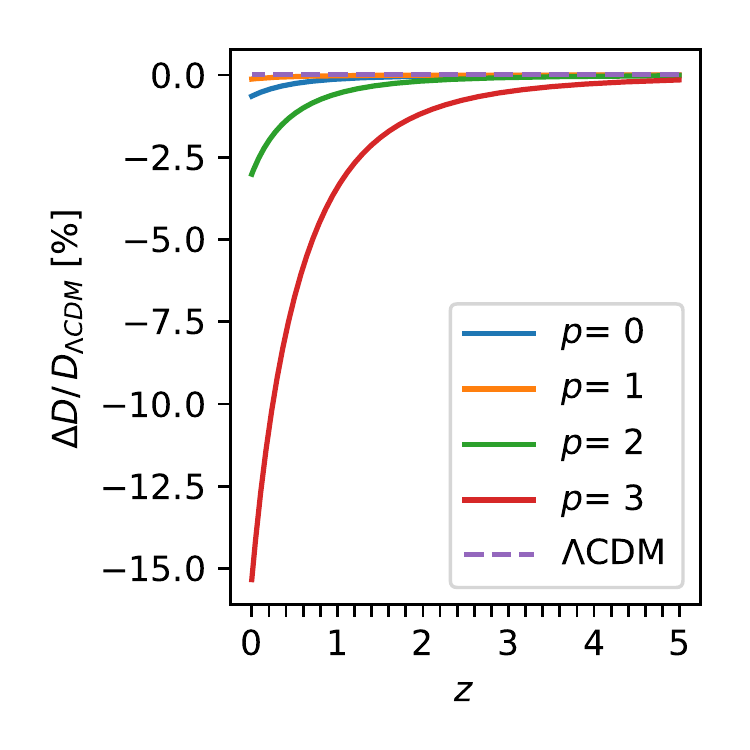}\hfill
  \includegraphics[width=0.333\textwidth]{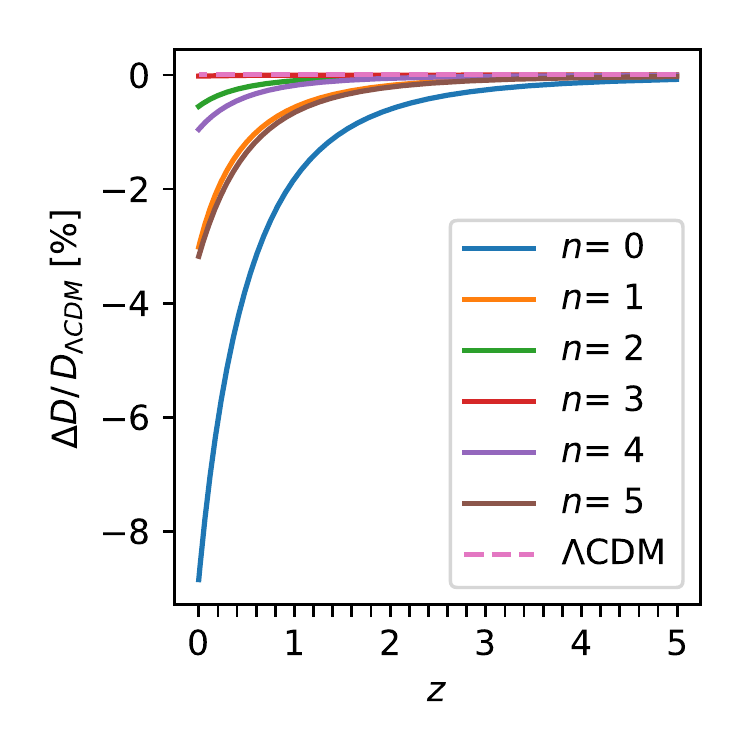}
  \caption{The matter density perturbation growth factor $D$ is shown 
as the relative deviation from $\Lambda$CDM for
    different models varying $\alpha$, $p$ and $n$, using $\alpha=1$, $p=2$,
    $n=1$ and $\psi_{ini} = 1.5$ as the base model. The more the field moves, 
the more thawing, allowing dark energy to dominate earlier, suppressing 
matter growth. By contrast, 
the case with $n=3$ has the field frozen for $\psi_{ini}>1$ and so 
$w_0 \approx -1$
    (see the left panel of figure~\ref{fig:continuous-fini_n.png}). 
Higher $\psi_{ini}$ would 
freeze most models in the plateau, decreasing deviations from $\Lambda$CDM.}
  \label{fig:D}
\end{figure}

Similarly, the growth function of matter perturbations 
(figure~\ref{fig:D}) does not show a
significant change with respect to $\Lambda$CDM. 
Since the dark energy equation of state only deviates appreciably from 
$-1$ at low redshift, the growth factor is close to that of $\Lambda$CDM 
until recently. Only those models that roll significantly, falling off 
the plateau show more than percent level deviations. 

\begin{figure}[t]
  \centering
  \includegraphics[width=0.333\textwidth]{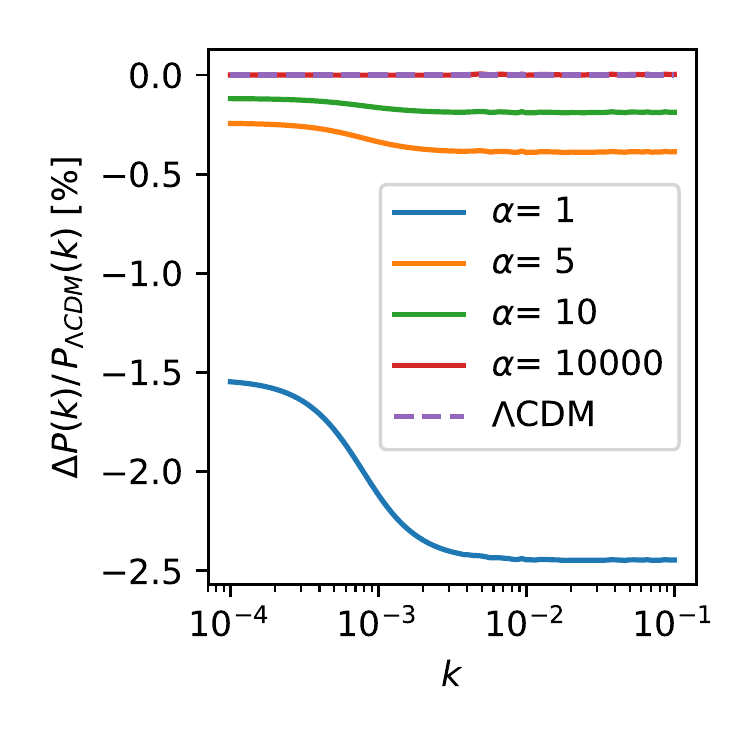}\hfill
  \includegraphics[width=0.333\textwidth]{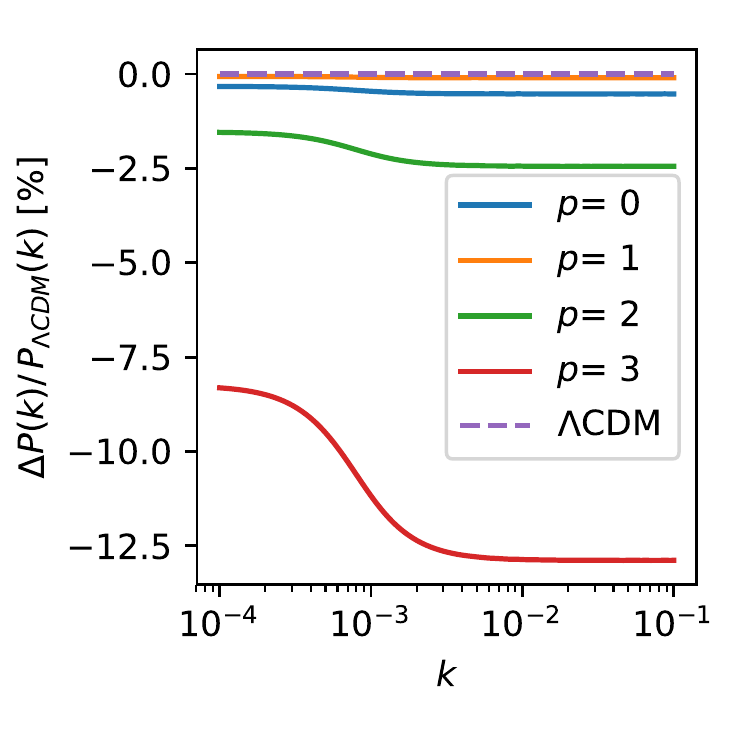}\hfill
  \includegraphics[width=0.333\textwidth]{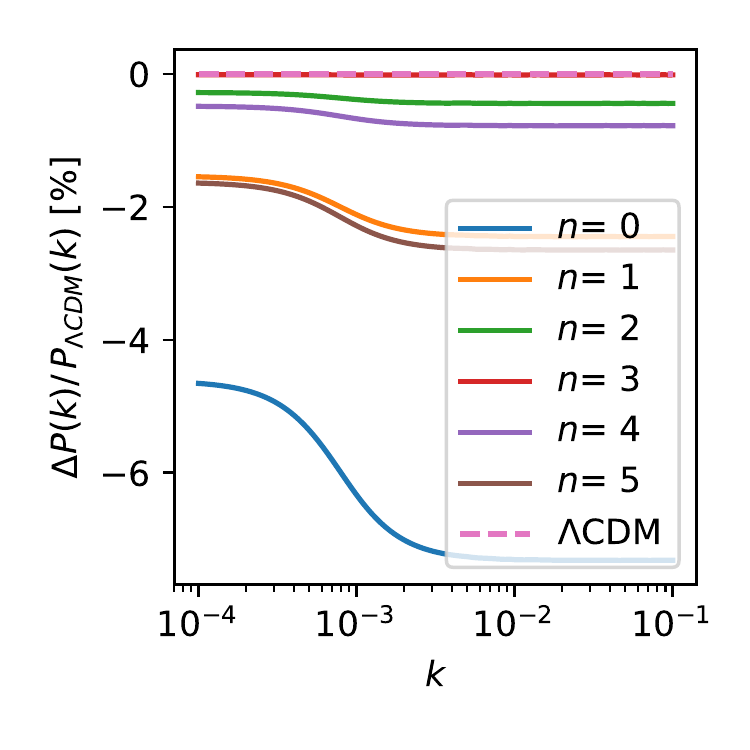}
  \caption{The matter power spectrum $P(k)$ is shown as the 
relative deviation from $\Lambda$CDM for
    different models varying $\alpha$, $p$ and $n$, using $\alpha=1$, $p=2$,
    $n=1$ and $\psi_{ini} = 1.5$ as the base model. The more the field thaws, 
the more the power spectrum is suppressed. 
} 
  \label{fig:pk}
\end{figure}

\begin{figure}[t]
  \centering
  \includegraphics[width=0.333\textwidth]{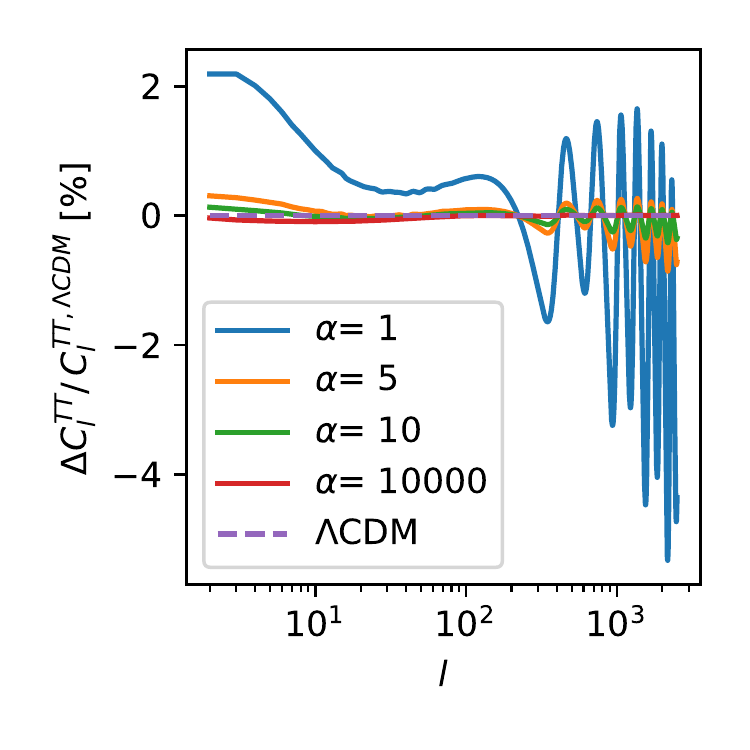}\hfill
  \includegraphics[width=0.333\textwidth]{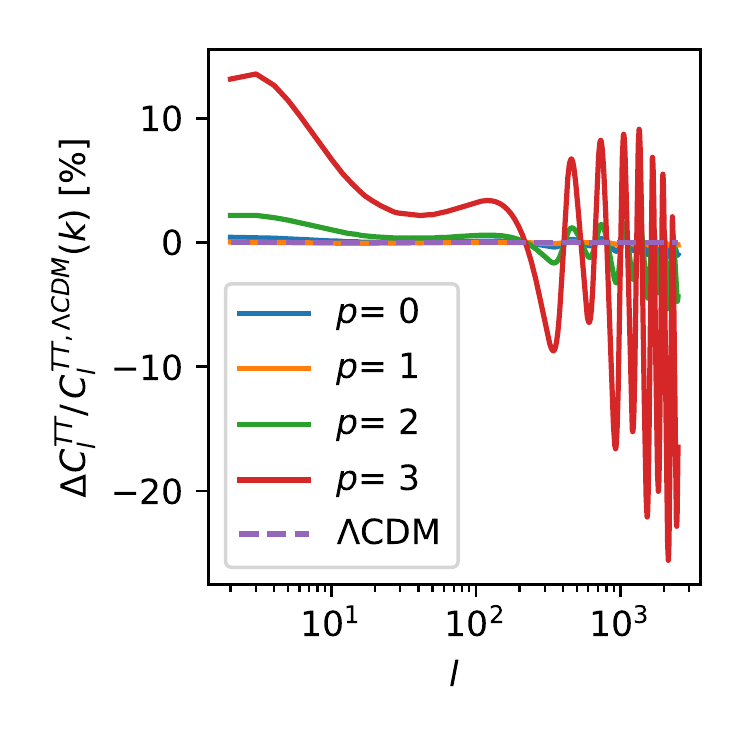}\hfill
  \includegraphics[width=0.333\textwidth]{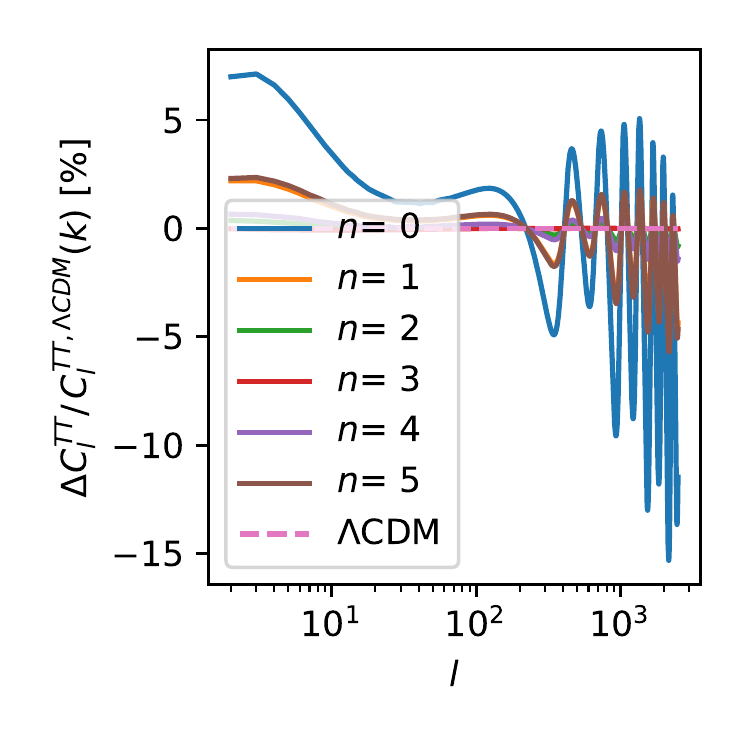}
  \caption{Temperature angular power spectrum ($C_l^{TT}$) 
is shown as the relative deviation
  from $\Lambda$CDM for different models varying $\alpha$, $p$ and $n$, using
  $\alpha=1$, $p=2$, $n=1$ and $\psi_{ini} = 1.5$ as the base model. 
Apart from having
  different amplitude for low $l$, the spectrum is shifted in angular scale with respect
  to that of 
$\Lambda$CDM.}
  \label{fig:cl}
\end{figure}

In those cases, the matter power spectrum as shown in 
figure~\ref{fig:pk} is lowered as well, yielding relative differences up to
a few percent. Similarly the CMB temperature angular power
spectrum (figure~\ref{fig:cl}) departs from $\Lambda$CDM. 
Such deviations can be compared with experimental data letting us 
rule some models out. In particular, the larger differences at high multipoles
are due to the geometric shift in the distance to recombination, anticipating 
that CMB and BAO galaxy distances will be important to take into account. 

\begin{figure}[t]
  \centering
  \includegraphics[width=0.333\textwidth]{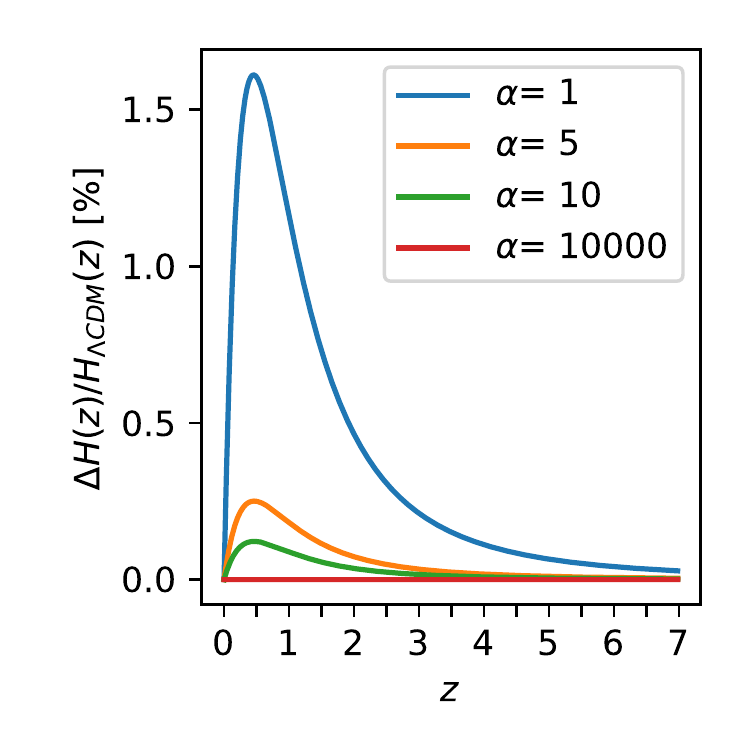}\hfill
  \includegraphics[width=0.333\textwidth]{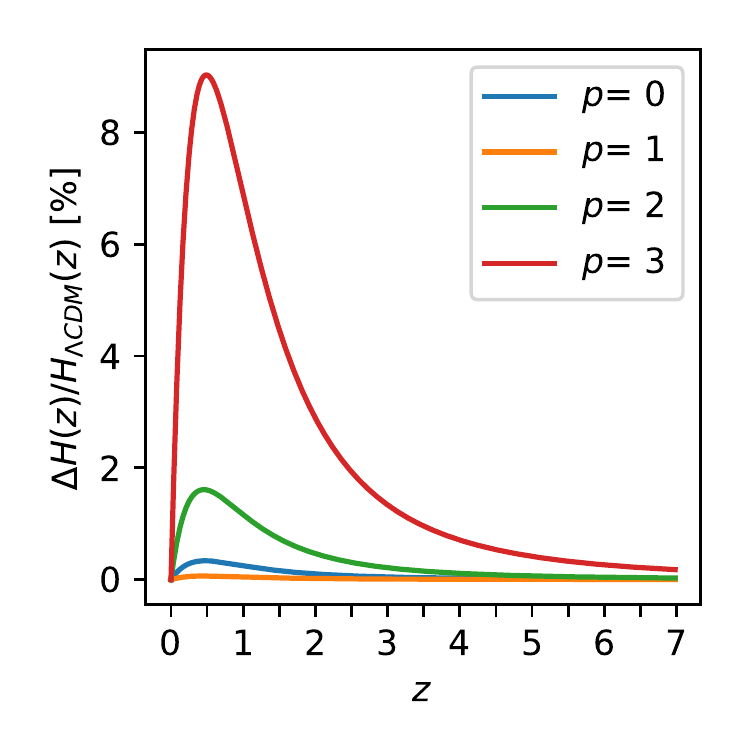}\hfill
  \includegraphics[width=0.333\textwidth]{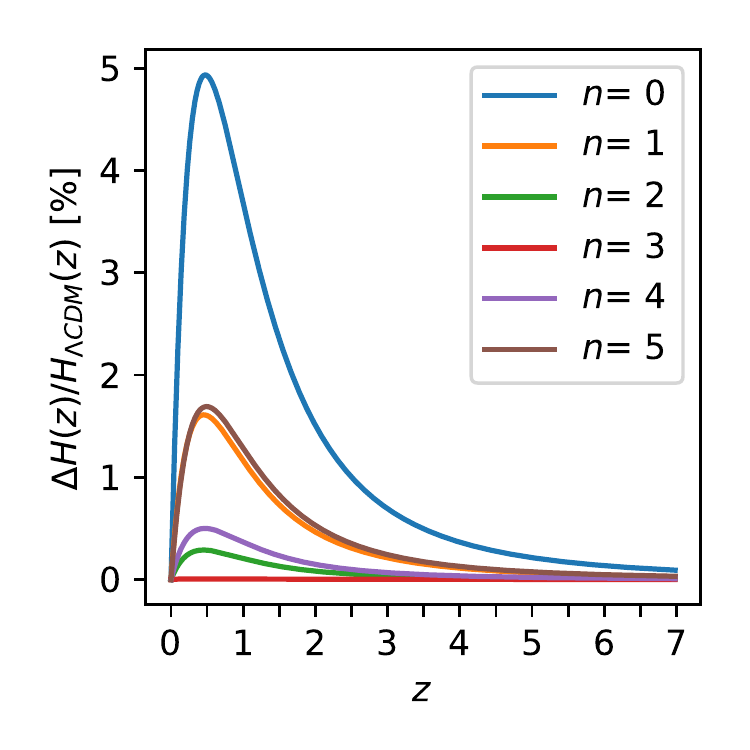}
  \caption{Hubble parameter is shown as the relative deviation from $\Lambda$CDM for
    different models varying $\alpha$, $p$ and $n$, using $\alpha=1$, $p=2$,
    $n=1$ and $\psi_{ini} = 1.5$ as the base model. Fields that thaw more 
have larger expansion rates, modulo the normalization to fixed $H_0$ today.}
  \label{fig:H}
\end{figure}

\begin{figure}[t]
  \centering
  \includegraphics[width=0.333\textwidth]{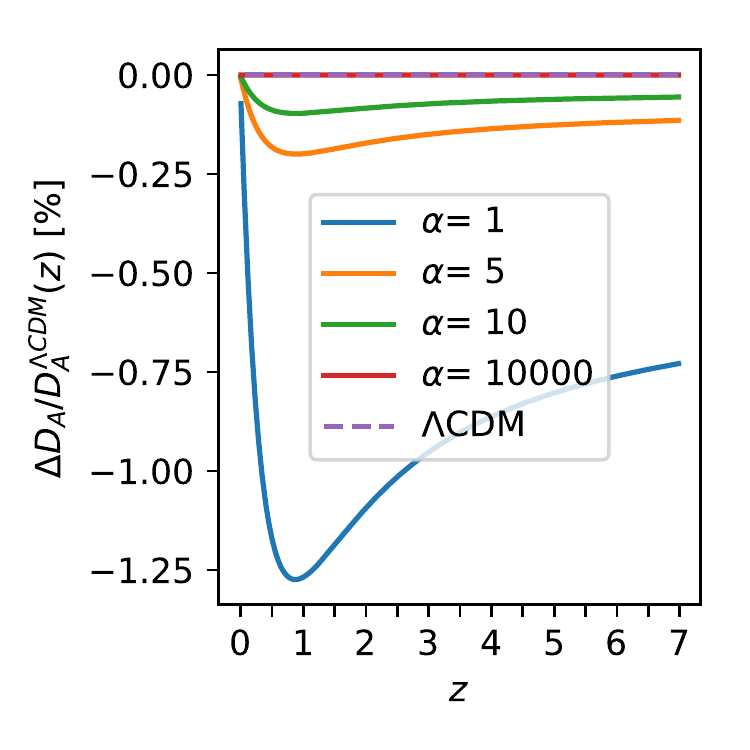}\hfill
  \includegraphics[width=0.333\textwidth]{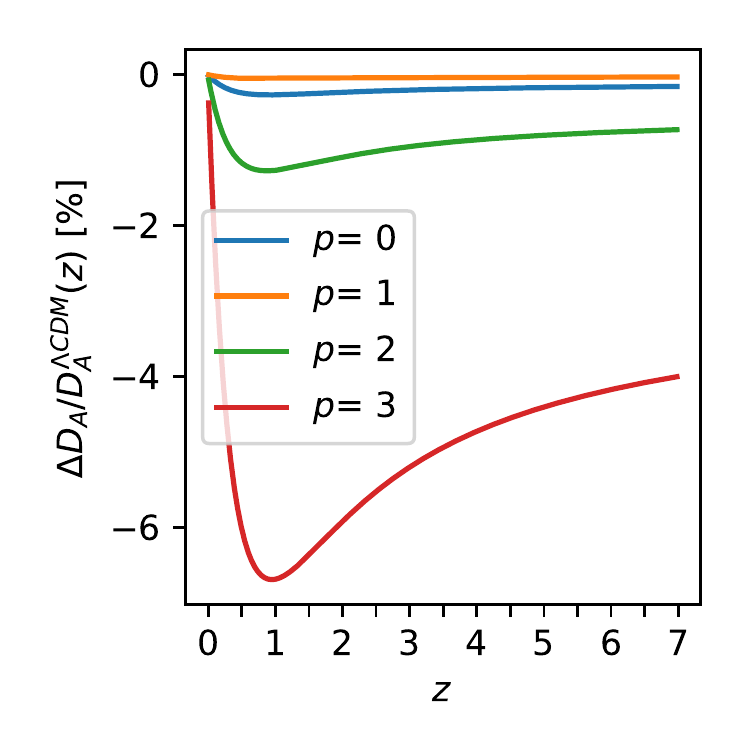}\hfill
  \includegraphics[width=0.333\textwidth]{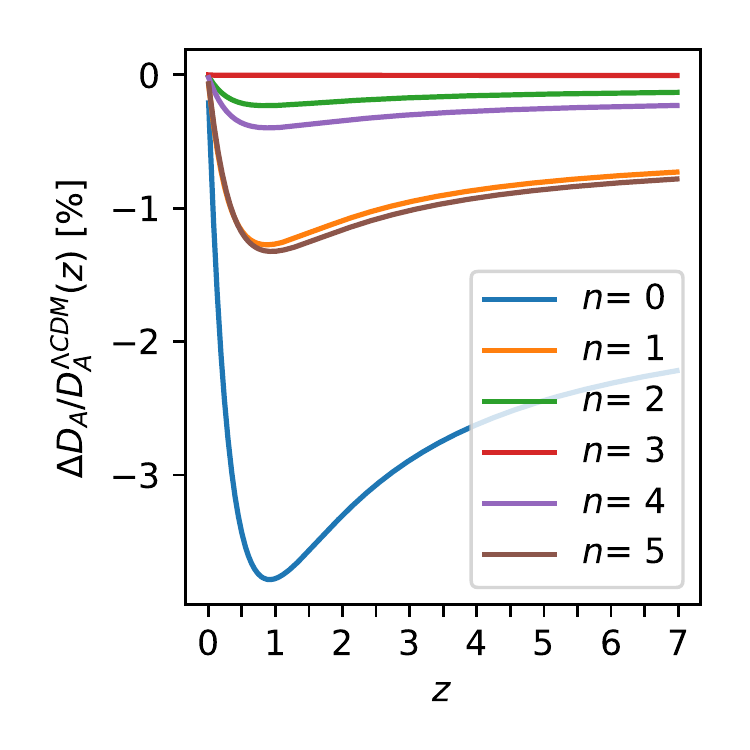}
  \caption{Angular distance is shown as the 
relative deviation from $\Lambda$CDM for
    different models varying $\alpha$, $p$ and $n$, using $\alpha=1$, $p=2$,
    $n=1$ and $\psi_{ini} = 1.5$ as the base model. 
} 
  \label{fig:DM}
\end{figure}

Distances are integrals over the Hubble parameter, i.e.\ 
\begin{equation} 
  D_A = \frac{1}{1+z} \int_0^z \frac{dz'}{H(z')} \,, 
  \label{eq:DA}
\end{equation} 
for the flat universe we consider. While the Hubble parameter should 
have its largest deviation from from $\Lambda$CDM at low redshift, as an 
integral the distance has increasing deviation with redshift. These 
quantities are shown in figure~\ref{fig:H} and figure~\ref{fig:DM}. 
Thus we expect that both $z<3$ measures, e.g.\ from supernovae and BAO, 
and high redshift measures from the CMB, will play important roles in 
constraints.

\section{Observational constraints}
\label{S:MCMC}

In this section we will compare the $\alpha$-attractor model (equation
\ref{eq:V}) with recent observations. In section~\ref{S:data} we will
explain the choice of datasets based on the knowledge acquired in previous
sections. Then, in section \ref{S:priors}, we will impose some appropriate and
sensible priors which will be needed in the study of the posterior
distributions of two iconic cases: that with the exponents of the Starobinsky
model ($p=2$, $n=1$) but with the scaling allowed to be free, 
and that with all the parameters freed. We will see that
a large portion of the parameter space is ruled out, favoring models 
close to $\Lambda$CDM.

\subsection{Data sets}
\label{S:data}

We will use the Planck 2015 observations \cite{Ade:2015rim}, BAO DR12
\cite{Alam:2016hwk} and the measurements of $E(z) = H(z)/H_0$, obtained
using the latest SN Ia at $z>1.5$ \cite{Riess:2017lxs, Scolnic:2017caz}.  The
reason we choose these observations is that Planck sets high redshift
constraints, although it also take into account the lower redshift effects
through integration, BAO does at low redshift, where we have found the largest
differences, and $E(z)$ imposes direct constraints in the expansion rate up to
$z = 1.5$. 

In general, CMB observations are used because CMB is affected by the expansion
rate, the matter components and inflationary conditions (a good summary of
reasons is found in section 2 of \cite{Ade:2015rim}), allowing the
constriction of the cosmological parameters. Indeed, for models close to
$\Lambda$CDM it allows precise evaluation of them. Furthermore, since
deviations from $\Lambda$CDM on the power spectrum rise up to few percents
(figures \ref{fig:pk} and \ref{fig:cl}), Planck and BAO likelihoods could be
sensitive to them. BAO set low redshift constraints on $H$
and $D_M = (1+z) D_A$, where the models' greatest differences with respect to
$\Lambda$CDM are found. We have chosen the latest released dataset, BAO DR12
\cite{Alam:2016hwk}, which covers $z\lesssim0.7$. 
We have not used Lyman-$\alpha$ BAO measurements,
even though they go much deeper in redshift, because they are in tension with
Planck and low redshift BAO measurements
\cite{Delubac:2014aqe,Evslin:2016gre}. For instance, in $\Lambda$CDM 
the discrepancy is approximately $2\sigma$, and our model is 
phenomenologically too close to $\Lambda$CDM to avoid the tension. 
Lastly, further control on
the expansion rate is given by the estimates of $E(z)$ up to
$z=1.5$ using the SN Ia distance observations at $z > 1.5$ \cite{Riess:2017lxs,
Scolnic:2017caz}. 

We used the full temperature, polarization and lensing power spectra from
Planck 2015~\cite{Ade:2015rim}.  The main signatures of the model are changes
on the cosmological background (at least for models close to a cosmological
constant) which can be tested by the Planck compressed likelihood (Table 4 in
Ref.~\cite{Ade:2015rim}). Including the full Planck data allows us to explore
potential degeneracies with other cosmological parameters
($A_s,\, n_s,\, \tau_{reio},\, \cdots$) and take into account effects in the
perturbation growth (i.e. ISW effect). The difference between the full and
compress Planck likelihoods is negligible for most of the common parameters,
with the largest deviation below 1$\sigma$ for $H_0$ and $\Omega_{cdm} h^2$, 
where $h = H_0/(\unit[100]{Mpc^{-1} km\, s^{-1}})$.

BAO DR12 measurements \cite{Alam:2016hwk} are independent of Planck's and are
in good agreement with them. Since the most deviating cases are ruled out from
the acoustic scale, we can just consider the background quantities and use the
consensus BAO-only values suggested in the paper. We have taken them from its
electronic
archive\footnote{\url{https://data.sdss.org/sas/dr12/boss/papers/clustering/ALAM_ET_AL_2016_consensus_and_individual_Gaussian_constraints.tar.gz}}.
Let us note that the actual variables that BAO can measure are $D_M(z)/r_d$
and $H(z)r_d$, which take into account changes in the cosmological parameters
and in the pre-recombination energy density era. Since $r_d$ is the sound
horizon at drag epoch ($z_d$), $r_d = \int_{z_d}^\infty dz\, c_s(z)/H(z)$,
with $c_s$ the speed of sound, and our models are nearly identical to
$\Lambda$CDM at high redshift, the BAO data informs us of deviations in $D_M=
(1+z) D_A$, i.e.\ the comoving distance and $H$. 

Finally, $E(z)$ constraints \cite{Riess:2017lxs} compress the
information of a thousand SN Ia distances from the unreleased Pantheon
dataset and Hubble Space Telescope, 
including 15 new discovered supernova at redshift $z>1.5$.
Their measurements are thought to be precise and unbiased estimates 
of $E(z)$ as
long as the expansion rate does not vary much nor have localized features
between the specific chosen redshifts, as for our model. Their only assumptions
were flatness, not mandatory but convenient, and continuity and smoothness of
$E(z)$ in order to parametrize its inverse by its values at different specific
redshifts. They used those points to interpolate and recover the whole
function, necessary to compute the luminosity distance (equation \ref{eq:DA}),
which they could compare against their selection of well-calibrated SN data.

\subsection{MCMC setup}
\label{S:priors}

The posterior distribution of the $\alpha$-attractor model (equation
\ref{eq:V}) which will be shown in next section has been obtained by 
sampling the parameter space with a Markov chain Monte Carlo (MCMC) 
method.  In particular, we made use of MontePython
\cite{Audren:2012wb} and CosmoHammer \cite{Akeret:2012ky} which embeds emcee
\cite{emcee}, an implementation of the Goodman and Weare affine invariant
ensemble sampler \cite{goodman2010ensemble}. We chose this algorithm instead
of the traditional Metropolis-Hasting to avoid acceptance rate and
convergence problems since affine invariant sampling methods are uniformly
effective over all the convex bodies with same space dimension and regardless
of their shape \cite{goodman2010ensemble}. In addition, CosmoHammer
parallelization allows much faster computations.

The priors are summarized in table \ref{tab:priors}.  Since the parameters 
for viable models do not vary over much more than an order of magnitude, 
and no particular values are preferred, a uniform prior is reasonable. 
This matches well with 
the MontePython implementation of CosmoHammer which does not allow 
informative or unbounded priors. 
As we have seen in section~\ref{S:Model}, sufficiently high
values of the scaling $\alpha$ or the initial field $\psi_{ini}$ all yield
$\Lambda$CDM-like universes.  Also, the exponents, $p$ and $n$, just under the
condition of being close to each other, no matter how high they were, yield
more $\Lambda$CDM-like results provided that the field had not started rolling
off the plateau too early, i.e.\ the viable set of models.  
There is therefore no point in allowing a large range of 
such high values, which all give essentially $\Lambda$CDM. 
Therefore we bound their space to prevent walkers\footnote{A walker is
  the equivalent of a chain in a Metropolis-Hasting algorithm, in
  Goodman-Weare terminology. Note their dynamics are different,
though.} wasting time in the infinite $\Lambda$CDM regime, although we 
set the bounds large
enough to be able to explore all the interesting phenomenology of the model.
Finally, in order to accelerate the convergence, we fixed $c$ with the closure
relation $1 = \sum_i \Omega_i$, using a bijection method.

\begin{table}
  \centering
  \begin{tabular}{|c|c|}
    \hline
    Parameter   &Range            \\
    \hline
    $H_0$       &[40, 90]         \\
    $\Omega_b h^2$  &[0, 0.04]    \\
    $\Omega_{cdm} h^2$ &[0, 0.3]  \\
    \hline
    $\dot \phi_{ini}$ &\{0\}               \\
    $\psi_{ini}$   &[0, 3.5]         \\
    $\alpha$    &[0, 10]          \\
    $p$         &[0, 10]          \\
    $n$         &[0, 10]          \\
    \hline
    $c$         &Fixed by $1 = \sum_i \Omega_i$ \\ 
    \hline
  \end{tabular}
  \caption{Priors used in the MCMC posterior inference. $H_0$ units
    are $\unit[]{[km\, s^{-1} Mpc^{-1}]}$ and $c$ is obtained by a bijection
  method.}
  \label{tab:priors}
\end{table}

\subsection{Starobinsky form vs $\Lambda$CDM}
\label{S:MCMC1}

Before studying the full general 
model posterior distribution, we want to focus on
the more constrained case with exponents $p=2$ and $n=1$, the one which
reduces to the Starobinsky potential when the scaling $\alpha=1$. 
Increasing the parameter space in stages will let us understand more easily 
how the model behaves, letting us check our understanding of its
dependence on $\alpha$ and $\psi_{ini}$. Furthermore, the scaling acts as 
an interpolation between the 
$\Lambda$CDM (high $\psi_{ini}$ or $\alpha$) and Starobinsky models
($\alpha=1$). 

After studying the autocorrelation times, we decided to use  $10$ walkers per
varied parameter, i.e.\ 230 walkers in total, and run 1400 iterations, from
which we discarded the first 1000 as burn in. Fewer iterations would have
sufficed since the slowest autocorrelation function crosses $0$ around
$400$, but we took a conservative approach given the low time cost. To
analyze convergence we used the MontePython \cite{Audren:2012wb} internal 
routine which puts all chains together, ordered by iteration step, splits the
resulting chain in three and calculates the Gelman-Rubin test. 

\begin{figure}[t]
  \centering
  \includegraphics[width=\textwidth]{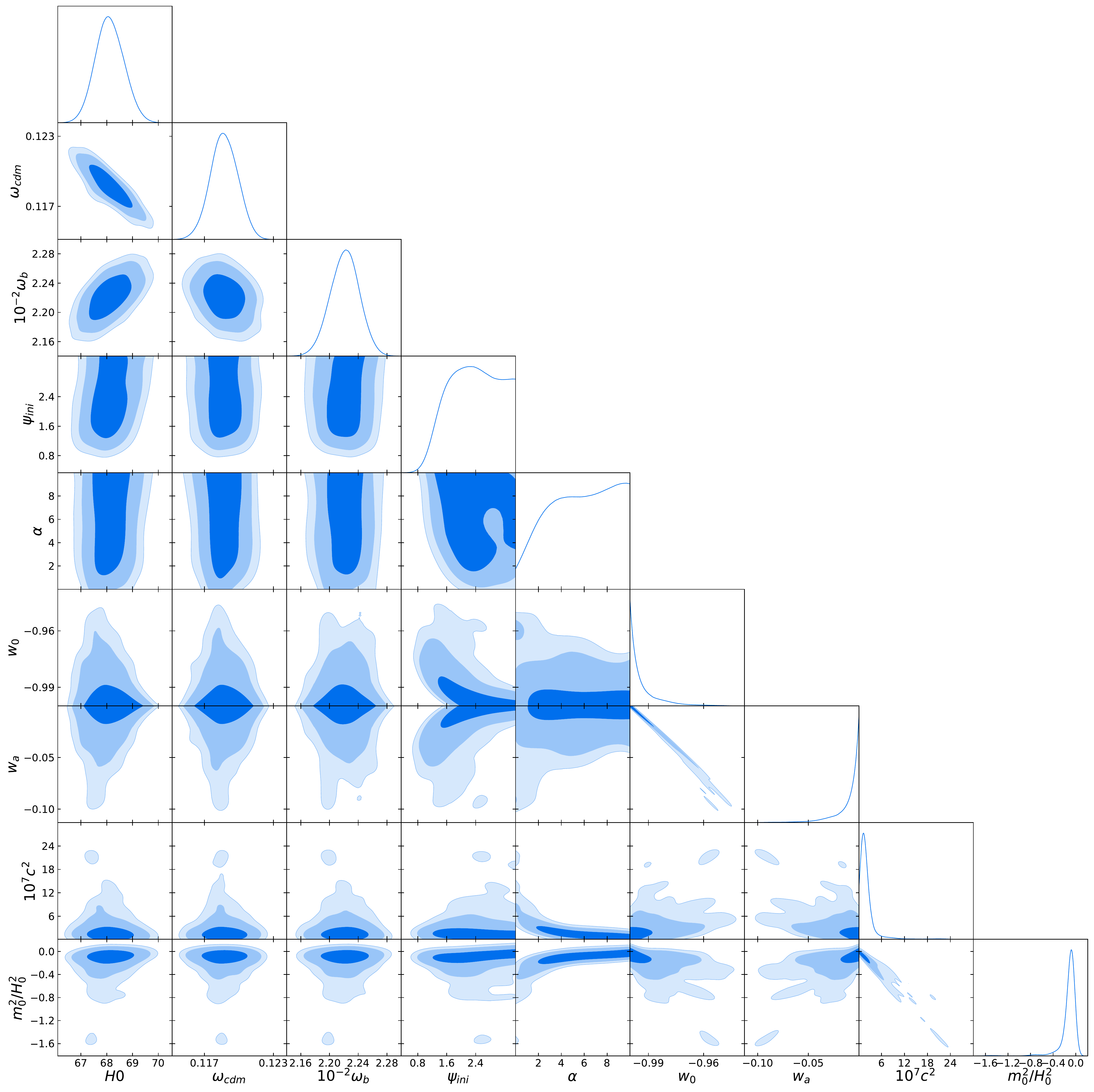}
  \caption{Posterior distributions for the Starobinsky form 
allowing $\alpha$ to vary, with $p=2$, $n=1$. The
    contours show the 68.3\%, 95.5\%, and 99.7\% confidence levels, with
    darker colors for more probable results. The quantitative results are
  summarized in table \ref{tab:Starobinsky-bf}.} 
  \label{fig:Starobinsky-posterior}
\end{figure}

The posterior distributions are shown in figure
\ref{fig:Starobinsky-posterior} while the quantitative results are in table
\ref{tab:Starobinsky-bf}. The first thing to note is that the cosmological
parameters are in good agreement with $\Lambda$CDM Planck 2015 results
\cite{Ade:2015xua}. This is related to the fact that our model has a
preferred $\Lambda$CDM-like regime. In fact, we see that their posterior
distribution is unaffected by $\alpha$ and almost any $\psi_{ini}$. The 
$\psi_{ini}$
posterior distribution, however, has a lower boundary which sets $\psi_{ini} >
1.27$ at 95.4\% confidence level. This value is altered by the parameters range
choice, as we will see in next section, and cannot be understood as a
general condition. The $\alpha$-$\psi_{ini}$ figure confirms what we said in
section \ref{S:alpha}: the lower $\alpha$ and $\psi_{ini}$ regime is disfavored
as a consequence of the fact that a faster evolution of the field yields, for
models that match the present dark energy density, a less negative equation 
of state, which is in tension with data. On the contrary, large $\alpha$ and
$\psi_{ini}$ freeze the field evolution getting it closer to $w = -1$, which 
the data prefer.

The dark energy equation of state parameters $w_0$ and $w_a$ confirm our
previous comments. Their values are close to $w=-1$, regardless of the
particular value of $\alpha$ and $\psi_{ini}$. Again, the reason why this 
happens is that a large value of either one of them can give $w\approx-1$ (see
for instance the $\alpha=10^4$ case in figure
\ref{fig:continuous-fini_alpha-prop}). 

\begin{figure}[t]
  \centering
  \begin{subfigure}[t]{0.45\textwidth}
    \includegraphics[width=\textwidth]{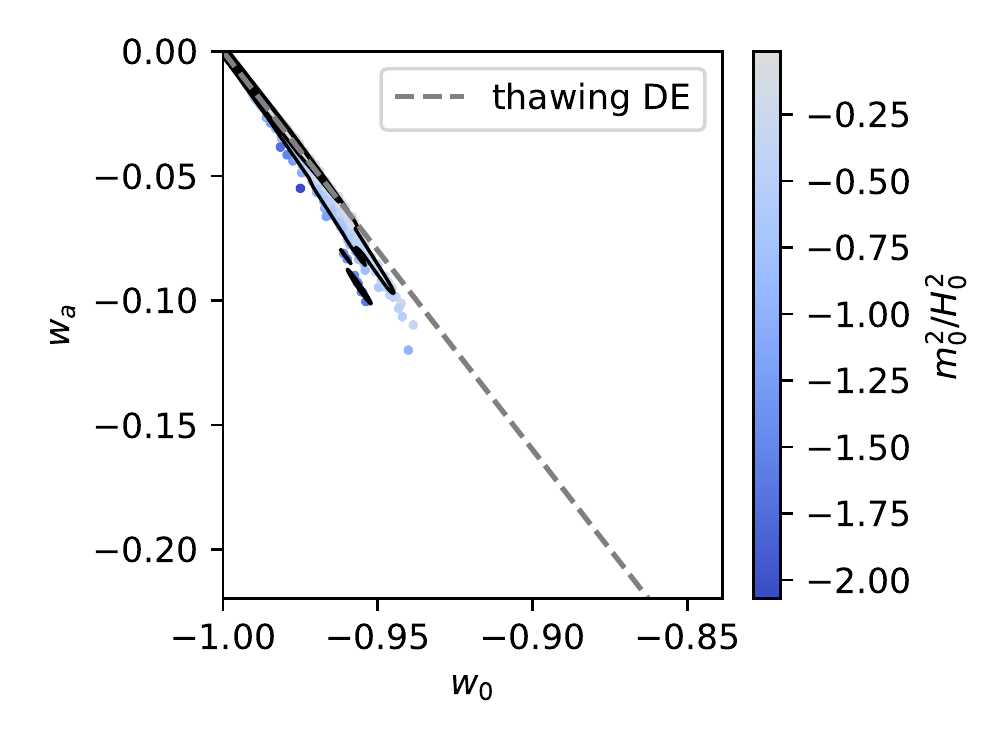}
    \caption{Starobinsky-like ($p=2$, $n=1$).}
      \label{fig:Starobinsky-w0wa}
    \end{subfigure}
    ~
    \begin{subfigure}[t]{0.45\textwidth}
      \includegraphics[width=\textwidth]{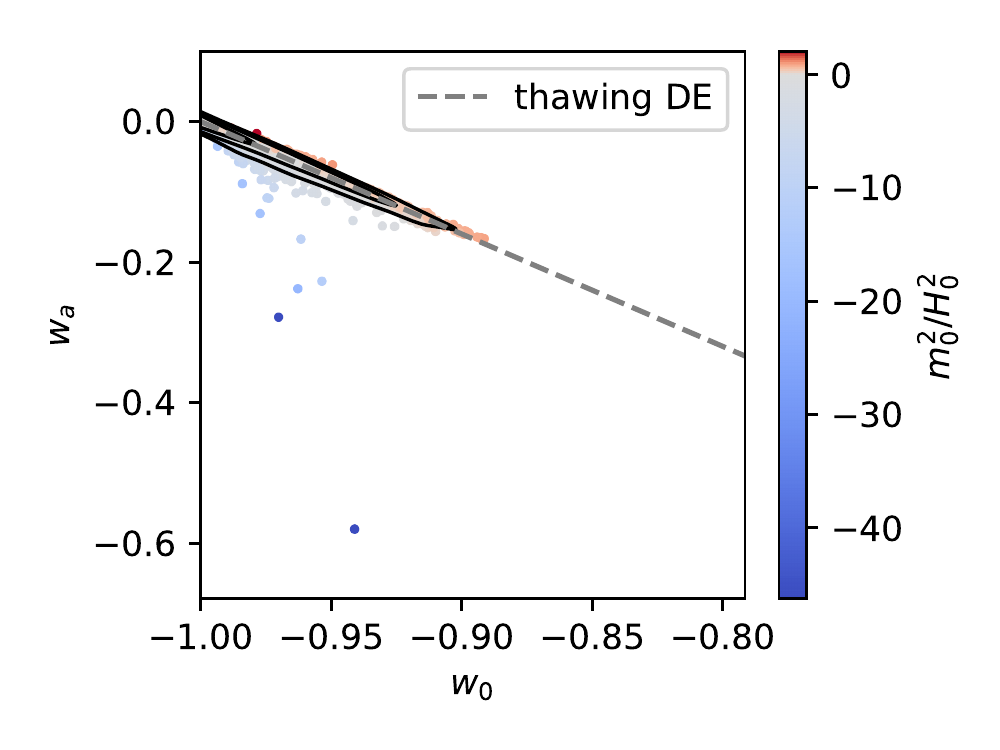}
      \caption{General.} 
      \label{fig:full-w0wa}
    \end{subfigure}
  \caption{The $w_0$--$w_a$ joint posterior distribution is shown for the 
Starobinsky form (left) and full generalized $\alpha$-attractor (right) 
      dark energy models (equation \ref{eq:V}) over a random sample of
      $2\times10^{4}$ points colored depending on their present field mass.
      The dashed grey diagonal line represents the usual $w_0-w_a$ relation 
for thawing models
      ($w_0<-0.8$) described in reference \cite{Linder:2015zxa}.
    }
  \label{fig:w0wa}
\end{figure}

The correlation between $w_0$ and $w_a$ follows the typical pattern of 
thawing fields. 
In figure
\ref{fig:Starobinsky-w0wa}, we have plotted the $w_0-w_a$ posterior
distribution over a random sample of $2\times 10^{4}$ points colored as a 
function of the value of the present field mass. Those models with 
greatest probability follow the thawing relation $w_a\approx -1.6(1+w_0)$ 
\cite{Linder:2015zxa}. This leads to tight constraints on $w_a$. 
A more subtle effect is the 
correlation between the equation of state parameters and the field mass, which
makes lower mass squared field models be located under those with higher 
mass squared for the same $w_0$. Recall that negative mass squared tends 
to occur for low $\alpha$ models, which also have greater deviations from 
$w=-1$. 

The amplitude parameter $c$, which we have obtained by requiring $1 = \sum_i
\Omega_i$, does not vary much; basically to get $\Omega_{de,0}\sim1$, 
one requires $c^2\lesssim H_0^2$, hence 
$c^2 \sim \unit[10^{-7}]{Mpc^{-2}}$. This will change once we let $n$ vary 
since $n$ affects 
the height of the plateau and $c$ must compensate any change on
the potential to obtain the correct dark energy density.  Finally, the present
field mass squared, $m_0^2/H_0^2$, is preferably negative, with 95\% CL
  region $(m_0^2/H_0^2)_{95\%CL} \in (-0.37, 0.07)$. The reason why most
  solutions are tachyonic is the absence of a maximum in the potencial. In
  this configuration, the regime with moderate slope, where the field can
  slow-roll, is found close to the plateau and has $V'' < 0$. However, the
  $V''>0$ regime is that described by a power law and, consequently, the field
  is quickly speed up. In spite of tachyonic solutions being natural they are
not large enough as to play an observable role, given the small effect of
even $m^2/H_0^2 \approx -20$ in section \ref{S:obs}. 

Before moving to full model we note that the best-fit Starobinsky form 
model has $\alpha=3.78$, well inside the 68\% confidence level region,
in contrast to the Starobinsky model ($\alpha=1$), which lies on the 95\% CL
region. In fact, although the $\chi^2_{\rm min}$ difference is relatively
small, for one more parameter, the Akaike Information Criterion
\cite{akaike1974new}, $\mbox{AIC} = \chi^2 + 2k$, where $k$ is the number of
model parameters, tells us that the Starobinsky case is disfavored over the
general one ($\mbox{AIC}_{BF} - \mbox{AIC}_{\alpha=1\,, BF} = -6.7$). The
Shchwartz information criterion ($\mbox{BIC} = \chi^2 + k \log(N)$, where $N$
is the number of data degrees of freedom) \cite{Schwarz:1978tpv}, however, is
more lenient over the $\alpha=1$ case as it takes into account the size of the
data sample. In this case, $|\Delta\mbox{BIC}| \sim 10^{-1}$, which does not
select any model as preferred. It is important to remark that, in both cases,
one finds that data prefers $\Lambda$CDM-like cosmologies, described by their
equations of state which are almost $w = -1$ ($w_0 = 0.998\,, 0.995$,
respectively). 

\begin{table}[t]
  \centering
  \begin{tabular}{|l|c|c|c|}
    \hline 
 Parameter                   &best-fit    &best-fit with $\alpha=1$  &  95\% limits\\
\hline                                     
{$\Omega_{cdm } h^2$}        & $0.1199 $  & $0.1192$    & $0.1187 \pm 0.0023$\\
{$H_0            $}          & $67.8 $   & $68.0$       & $68.1^{+1.1}_{-1.0}     $\\
{$10^{2}\Omega_{b } h^2$}    & $2.216 $   & $2.226$     & $2.221^{+0.039}_{-0.040}   $\\
$10^{-9}A_s$                 & $2.12 $   & $2.12$       & $2.14^{+0.10}_{-0.090} $\\
$n_s$                        & $0.9592 $  & $0.9652$    & $0.9644^{+0.0088}_{-0.0089}$\\
$\tau_{reio}$                & $0.063 $   & $0.058$     & $0.068^{+0.024}_{-0.023}   $\\
\hline                                     
{$\psi_{ini}        $}       & $2.61 $    & $2.82$    & $> 1.27                    $\\
{$\alpha         $}          & $3.78 $    & $1.00$    & ---                         \\
$10^{7} c^2$                 & $1.45 $     & $5.32 $    & $ < 5.5      $\\
\hline                                     
{$w_0            $}          & $-0.998 $  & $-0.995$    & $-1\leq w_0<-0.974              $\\
{$w_a            $}          & $-0.0029 $  & $-0.0081$    & $0 > w_a > -0.0413  $\\
$m_0^2/H_0^2$                & $-0.09 $  & $-0.28$    & $-0.12^{+0.19}_{-0.25}    $\\
\hline
$\chi_{min}^2/2$             & $5642.099 $  & $5646.458$    & --- \\
\hline
 \end{tabular}
 \caption{Parameter best fit values and 95\% confidence limit for the 
$p=2$, $n=1$ Starobinsky form, allowing $\alpha$ to vary. The 
$\alpha = 1$ column corresponds
   to the pure Starobinsky model. 
$H_0$ units are $\unit[]{km\, s^{-1} Mpc^{-1}}$ and
   $\unit[]{Mpc^{-2}}$ for $c^2$. Note that $\alpha$ is unbounded 
 at 95\% CL (see figure \ref{fig:Starobinsky-posterior}).}
 \label{tab:Starobinsky-bf}
\end{table}

\subsection{Analysis of the full posterior distribution}
\label{S:MCMC2}

Finally, we will study the general case with free exponents $p$ and $n$, 
as well as $\alpha$.
It will collect all the dependencies we have been describing and let us
find the full posterior distribution for the dark energy model proposed by
Ref.~\cite{Linder:2015qxa}, an $\alpha$-attractor quintessential model with
potential given by equation \ref{eq:V}. 

For this case the number of walkers per parameter was maintained in 10,
giving a total of 250 walkers, the sample iterations were increased to 2500,
with 1000 burn in iterations to be discarded, and 1500 iterations to be
stored.  The extra degrees of freedom had a direct impact on convergence,
making it much slower.  As before, we checked the autocorrelation time and
computed the Gelman-Rubin test with MontePython \cite{Audren:2012wb}. 

\begin{figure}[t]
  \centering
  \includegraphics[width=0.75\textwidth]{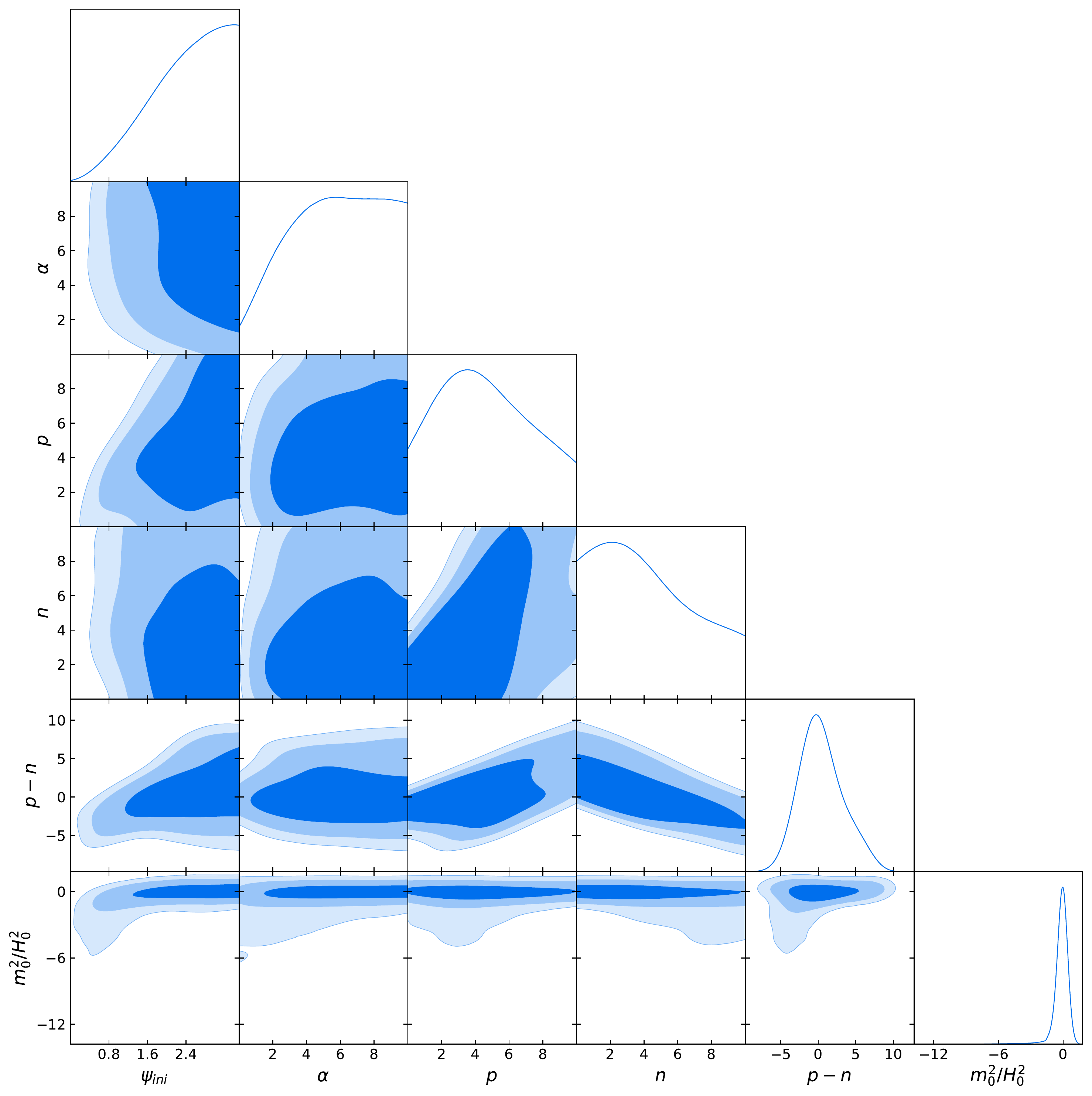}
  \caption{Posterior distribution for the generalized 
$\alpha$-attractor dark energy model
    with potential \ref{eq:V}. The contours show the confidence levels up to
    99.7\% CL, with darker colors for more probable results. The quantitative
    results are summarized in table \ref{tab:Full-bf}.} 
  \label{fig:Full-posterior-short}
\end{figure}

The results are written in table \ref{tab:Full-bf}. In figure
\ref{fig:Full-posterior-short} we have plotted the marginalized posterior
distribution of the model parameters and the present mass field, leaving aside
the cosmological parameters because of their similarities with the results in
previous section, represented in figure \ref{fig:Starobinsky-posterior}. 
Note that the $\alpha-\psi_{ini}$ plot continues to show 
that $\psi_{ini}$ and the kinetic coefficient $\alpha$ cannot be
simultaneously small.  Apart from that, all the parameter subspaces 
tell us something new. For example, the high exponent $p$ disfavors the lower
$\psi_{ini}$ regime. As we said in section \ref{S:pn}, higher $p$ makes the
low $\psi_{ini}$ regime steeper, making the field roll faster and changing the
expansion rate too much to match the observations.  Higher $n$, changing more
the amplitude of the plateau, do not have as strong an effect on the
$\psi_{ini}$ range, but does allow for much lower $c$; note that $n\sim5$ in
the middle of its range can suppress $c^2$ by a factor $2^{-2n}\sim 1000$. 

However, we see that actually it is the difference $p-n$ that 
mostly matters (see equation~\ref{eq:Vfgtr} and section~\ref{S:pn}), 
with a strong correlation in the $p$ vs $n$ plane. 
The mean posterior, with the 95.4\% CL values, is $p-n =
-0.4^{+6.4}_{-5.2}$.  Of course, higher values of $\psi_{ini}$ and $\alpha$
make the constraint on $p-n$ weaker as they freeze the field so there is less
sensitivity to the shape of the potential.  

The $m_0^2/H_0^2$ distribution is also slightly broader, thanks to the
  possibility of having a maximum which bounds the region where $V''<0$, and
  makes $V''>0$ at the plateau. As we know that slow-roll solutions are mainly
  placed in those regimes, the 95\% CL is symmetric around $0$, 
$|m_0^2/H_0^2|_{95\% CL} \sim 1 $.  Similarly, freeing the exponents allows the
$w_0-w_a$ distribution (figure \ref{fig:w0wa}) to expand, while continues
following the thawing solution, even better than before since allowing
$p\approx n$ can give flatter plateaus. 

It is important to note that introducing two new parameters has just
slightly improved the $\chi^2$ ($\chi^2_{full}/2 = 5641.276 \simeq 5642.098 =
\chi^2_{np\,, fixed}/2$).  This makes the Akaike Information Criterion
($|\Delta\mbox{AIC}| = 2.4$) prefer the simpler model with fixed exponents.
The Schwartz criterion is more strict in this case ($|\Delta\mbox{BIC}| ~
O(10)$), completely rejecting the more complex case. This result is a
consequence of the limited phenomenology available in this model, which is
mainly described by how much and how fast the field changes, which can be
controlled by just $\alpha$ and $\psi_{ini}$. Furthermore, the data prefers
$\Lambda$CDM which is recovered by high $\alpha$ or $\psi_{ini}$.  Indeed, the
best fit model is almost indistinguishable from $\Lambda$CDM, having $(w_0,
w_a) = (-1.000, -0.0006)$.

\begin{table}[t]
  \centering
  \begin{tabular}{|l|c|c|}
    \hline 
Parameter &  best-fit   & mean $\pm$ 95\% limits\\
\hline
{$\Omega_{cdm } h^2$}      & $0.1186$      & $0.1183^{+0.0024}_{-0.0022}$\\
{$H_0            $}        & $68.3$       & $68.2^{+1.0}_{-1.1}$\\
{$10^{2}\Omega_{b } h^2$}  & $2.224$       & $2.221 \pm 0.038$\\
$10^{-9}A_s$               & $2.15 $       &$2.14^{+0.10}_{-0.098}$\\
$n_s$                      & $0.9640 $     & $0.9649^{+0.0081}_{-0.0087}$\\
$\tau_{reio}$              & $0.070 $      & $0.067^{+0.026}_{-0.025} $\\
\hline                           
{$\psi_{ini}        $}     & $1.401$       & $> 0.955                   $\\
{$\alpha         $}        & $8.530$       & ---                         \\
$10^{3} c^2        $       & $3.44\times 10^{-3}$         & $ < 3.89         $\\
{$p              $}        & $3.140$       & ---                         \\
{$n              $}        & $4.233$       & ---                         \\
$p-n$                      & $-1.1$        & $0.4^{+6.4}_{-5.2}        $\\
\hline                           
{$w_0            $}        & $-1.000$      & $-1\leq w_0< -0.951$\\
{$w_a            $}        & $-0.0006$      & $0 > w_a > -0.0789$\\
$m_0^2/H_0^2$                & $-0.19$       & $-0.2^{+1.2}_{-1.1}$\\

    \hline
    \multicolumn{3}{|c|}{$-\ln{\cal L}_\mathrm{min} = 5641.276$}\\
    \hline 
 \end{tabular}
 \caption{Best fit model for the full potential \ref{eq:V}. $H_0$ units are
 $\unit[]{km\, s^{-1} Mpc^{-1}}$ and $\unit[]{Mpc^{-2}}$ for $c^2$. Note that
 $\alpha$, $p$ and $n$ are unbounded (see figure
 \ref{fig:Full-posterior-short}). 
} 
 \label{tab:Full-bf}
\end{table}

\subsection{Future observational data sets} 
\label{S:mission} 

Future experiments will be able to test in more detail 
the expansion histories of the $\alpha$-attractors models studied in this 
paper that differ more subtly from the $\Lambda$CDM cosmological 
constant predictions. Using BAO 
measurements, we will have additional empirical information on
the Hubble parameter $H(z)$ and the angular 
distance determinations $D_{A}$. From SN~Ia we 
will obtain distance luminosity determinations 
$D_{L}$  at various redshifts. 

The Dark Energy Spectroscopic Instrument (DESI) will be a highly 
constraining BAO experiment. 
DESI will use four target classes from redshift z $>$ 0.5 to z $<$ 4
\cite{Aghamousa:2016zmz}. Those include Luminous Red Galaxies (LRGs), 
Emission Line Galaxies (ELGs), QSO and Lyman--alpha QSOs. In addition
to the BAO measurements, this experiment will be able to discriminate 
models through growth rate of structure, with a precision comparable 
to that from weak lensing, through the redshift--space--distortions 
(RSD) method, which requires a very good sampling of the large scale 
structure. The Euclid satellite will also give highly accurate measurements 
for $z\gtrsim1$ \cite{Amendola:2016saw}. 

The WFIRST satellite will measure 
the expansion history of the Universe using 
supernovae. WFIRST will obtain thousands of well
observed SN Ia and it is possible that it will observe SNe Ia identified 
by other telescopes such as the Large Synoptic Survey
Telescope (LSST) \cite{Hounsell:2017ejq}. 
The growth of structure of the Universe will be also obtained by WFIRST. 

These are all very 
powerful cosmological tools to discriminate between the candidates studied
here against other physical models of dark energy.

\section{Conclusion}
\label{S:conclusion}

We have studied the $\alpha$-attractor dark energy model
proposed in Ref.~\cite{Linder:2015qxa}, and expressed in equation~\ref{eq:V},
inspired by the $\alpha$-attractor class of inflationary models. It is a
generalization of the well known Starobinsky potential which correspond to the
case with kinetic coefficient $\alpha=1$ and exponents $p=2$ and $n=1$, with no
coupling to matter since it is not obtained from a conformal transformation
but from gauge fixing the extra degree of freedom associated to the conformal
symmetry of the inflationary Lagrangian. This model is appealing because it is
able to interpolate between the two most common $\alpha$-attractor forms 
used for inflation -- the Starobinsky model and the T-model -- and also 
interpolate between power law potential and $\Lambda$CDM dark energy regimes 
at low and high $\psi$, respectively. This potential also allows, in theory,
the existence of clustering dark energy due to tachyonic solutions.

We can summarize the key points of this paper in the following ideas:
\begin{itemize}
  \item The model is mainly described by its background evolution, which
    depends on the parameters in the following way: 
    \begin{itemize}
      \item At viably high $\psi$, it is in the thawing class, with 
the field excursion related to the equation of state and the $\alpha$ 
parameter as $\Delta \psi \sim \sqrt{1+w_0}/\sqrt{\alpha}$ and 
$1+w_0\sim 1/\alpha$ so $\Delta \psi \sim 1/\alpha$. 
      \item If $p>n$, the field always decreases and its speed is also
        inversely related to $\psi_{ini}$. In addition, $p$ controls the
        steepness of the transition regime.
      \item If $p<n$, the field can grow towards an asymptotic 
de Sitter state at infinity if
        $\psi_{ini}>\psi_{max}$. 
Around the
        inflection points, or too close to $\psi=0$, the field evolution is
        fast. The exponent $n$ controls the height of the maximum, and 
the difference $p-n$ controls the characteristic scale of the uplifted 
exponential potential. 
    \end{itemize}
  \item  The MCMC analysis showed that the $\Lambda$CDM-like regime is 
favored by the combined data set of 
    Planck 2015 \cite{Ade:2015rim}, BAO DR12 \cite{Alam:2016hwk} and
    $E(z)$ estimation from supernovae \cite{Riess:2017lxs}:
    \begin{itemize}
      \item  Both the initial field $\psi_{ini}$ and $\alpha$ are 
pushed to larger values, where the field lingers on the plateau. 
      \item  The case where $p \sim n$ is also favored as it keeps the 
potential from being too steep (the deviation from a flat plateau becomes 
second order in the uplifted exponential potential). 
      \item  The tachyonic solutions compatible with the observations 
have a very mild instability that does not give appreciable dark energy 
clustering. They 
        are, however, as likely as non-tachyonic ones: $m_0^2/H_0^2 =
        -0.01^{+0.97}_{-0.87}$.
    \end{itemize}
\end{itemize}

We studied the properties of the model and its dependence on the different
parameters. We saw that a better variable to understand the field evolution
is $\psi=\phi/\sqrt{\alpha}$ instead of $\phi$. The kinetic coefficient 
$\alpha$ (which for inflation models is intimately tied to the geometry 
of the superconformal field theory space) scales the
field so that it determines at what value the field thaws. On the
other hand, the reason why one has to consider both $\psi$ and $\alpha$ is that
the latter appears explicitly in the potential (but not kinetic) energy, 
in the frame in which the kinetic term is canonical (equation~\ref{eq:L}), 
so that different values of $\alpha$ yield different universe
histories. In particular, we saw that the field evolution rate is inversely
related to $\alpha$ -- higher values of $\alpha$ slow
it down.  We showed how the initial position of the field 
$\psi_{ini}$ played an important role determining how the field would evolve
and had to be taken into account. For $p>n$ the field would always roll down
but its speed would also decrease as $\psi_{ini}$ grows, placing the field
further on the plateau. On the contrary, when $p<n$ the potential has a maximum
whose height and position is controlled by the relative size of $n$ and $p$.
In this scenario, the field could roll down towards $\psi=0$, where the
potential behaves like a power law potential $V \propto \phi^p$ (equation
\ref{eq:Vflow}) or towards $\psi=\infty$, a de Sitter attractor with $V
\propto 2^{-2n}$ (equation \ref{eq:Vfgtr}), with speed dependent on the
proximity to the inflection points, where the force felt by the field is
maximal. This correlates the expansion history with the 
field  mass squared, and with its sign (which 
depends on which side of the inflection point the field is at present). 
In particular we have seen
(figure \ref{fig:w0wa}) that the equation of state parameters $w_0$ and $w_a$
are ordered so that for a given $w_0$ the higher mass squared gives higher
$w_a$. 

In section \ref{S:obs} we briefly studied how the negative mass squared 
could affect the observables.  We saw that it would only have an effect if 
$|m^2|/H^2 \gtrsim k^2/(a^2 H^2) > 1$ (equation~\ref{eq:k_mass}), 
which greatly constrains the relevant modes, leaving just those of order $k\sim
\unit[10^{-3}]{Mpc^{-1}}$.  Numerically, we studied the most extreme case we
had obtained, that with $m^2/H_0^2 \sim -20$.  In this case, we saw the mass
term became relevant at $z\sim 3$ and that it was just today when it reaches 
the same order as the mode $k = \unit[10^{-3}]{Mpc^{-1}}$. Therefore, the
dark energy density perturbations are not appreciably influenced by the 
tachyonic instability, except possibly in the future. 

For a detailed analysis we constrained the parameter space with 
MCMC for two $\alpha$-attractor models. 
We compared our theoretical predictions with current datasets of 
CMB (Planck 2015 angular power spectrum, polarization and
lensing \cite{Ade:2015rim}), BAO (BOSS DR12 \cite{Alam:2016hwk}) and
Supernovae (Pantheon + Hubble Space Telescope compressed \cite{Riess:2017lxs,
Scolnic:2017caz}). We also discussed how to choose priors since as some
parameters get large the models become insensitive to them and
indistinguishable from $\Lambda$CDM. 

The results for the 
two models were discussed in sections~\ref{S:MCMC1} and \ref{S:MCMC2}. The
first case corresponds to the particular case where the potential is
Starobinsky-like but leaving free the $\alpha$ parameter; i.e.~fixing $p = 2$
and $n=1$. We saw that the preferred models are those closer to $\Lambda$CDM
and the best-fit is almost $w=-1$. Indeed the $w_0$--$w_a$ behavior closely 
follows that of thawing dark energy for $\psi_{ini}$ not too small, and this 
is bounded from below. Furthermore, the other cosmological parameters
are compatible with those from Planck 2015 for $\Lambda$CDM.

These behaviors hold for the full generalized $\alpha$-attractor model, 
with parameters $\alpha$, $p$, $n$. The preferred region continued to 
be that closer to $\Lambda$CDM, and a new way of attaining that was for 
$p$ and $n$ to be nearly the same. 
Even though the model had much more freedom, 
the $w_0-w_a$ behavior (figure~\ref{fig:full-w0wa}) still followed 
the thawing fit $w_a=-1.6(1+w_0)$ quite well. 

In closing, we would like to comment on possible future work based on this
model. In this work, we have just studied a quintessence cosmological model,
leaving aside the inflation epoch from which it is originally inspired and
justified the absence of coupling to matter, even in the Starobinsky
potential. Therefore, it is sensible to think of the next work as an
investigation of the joint predictions of this generalized model over the whole
history of the Universe, from inflation to late time cosmology.  This 
kind of work
would be in the spirit of the recent papers on the subject 
\cite{Dimopoulos:2017zvq, Dimopoulos:2017tud, Akrami:2017cir} and 
might show that this model is well suited to address both inflation and dark
energy from a common physical mechanism, linking two of the fundamental
problems in modern cosmology.

\acknowledgments{CGG would like to thank his UCM colleagues Jos\'e M.
  S\'anchez Vel\'azquez, H\'ector Villarrubia-Rojo and Miguel Aparicio Resco
  for useful discussions. We thank Renata Kallosh for interesting 
discussions. CGG and PRL are supported by AYA2015-67854-P from the Ministry of
Industry, Sciece and Innovation of Spain and the FEDER funds. CGG is supported
by the Spanish grant BES-2016-077038. EL is supported in part by the Energetic
Cosmos Laboratory and by the U.S. Department of Energy, Office of Science,
Office of High Energy Physics, under Award DE-SC-0007867 and contract no.
DE-AC02- 05CH11231. MZ is supported by the Marie Sklodowska-Curie Global
Fellowship Project NLO-CO.  This paper is based upon work from COST action
CA15117 (CANTATA), supported by COST (European Cooperation in Science and
Technology). }

\bibliography{./aatt-paper.bib}
\bibliographystyle{JHEP}
\end{document}